\documentclass[useAMS,usenatbib]{mn2e}
\usepackage{graphicx}
\usepackage{color}
\usepackage{amssymb}
\usepackage{amsmath}
\usepackage{subfigure}
\usepackage{rotating}
\usepackage{pdflscape}
\usepackage{bm} % This makes letters bold and italic as needed in mathmode
\usepackage{mathtools}
\usepackage[colorinlistoftodos]{todonotes}

\usepackage{comment}
\includecomment{details}
\graphicspath{{images/}}

\title[{Modulation of Galactic cosmic rays}]{{The Galactic cosmic ray intensity at the evolving Earth and young exoplanets}}
\author[Rodgers-Lee et al]{D. Rodgers-Lee$^{1}$\thanks{E-mail:
drodgers@tcd.ie}, A. A. Vidotto$^{1}$, A. M. Taylor$^{2}$, P. B. Rimmer$^{3,4,5}$, T. P. Downes$^{6}$  \\ 
$^{1}$ School of Physics, Trinity College Dublin, University of Dublin, College Green, Dublin 2, D02 PN40, Ireland \\
$^{2}$ DESY, D-15738 Zeuthen, Germany \\
$^{3}$ {Department of Earth Sciences, University of Cambridge, Downing St, Cambridge CB2 3EQ, United Kingdom}\\
$^{4}$ {Astrophysics Group Cavendish Laboratory, JJ Thomson Ave, Cambridge CB3 0HE, United Kingdom}\\
$^{5}$ {MRC Laboratory of Molecular Biology, Francis Crick Ave, Cambridge CB2 0QH, United Kingdom}\\
$^{6}$ Centre for Astrophysics \& Relativity, School of Mathematical Sciences, Dublin City University, Glasnevin, D09 W6Y4, Ireland
}

\begin{document}
\date{Accepted 2020 September 3. Received 2020 August 24; in original form 2020 March 16}
\pagerange{\pageref{firstpage}--\pageref{lastpage}} \pubyear{xxxx}
\maketitle

%\author[Rodgers-Lee, Vidotto, Taylor]{D. Rodgers-Lee$^{1}$\thanks{E-mail:
%drodgers@tcd.ie}, A. Vidotto $^{1}$, A. M. Taylor $^{2}$ \\ 
%$^{1}$ School of Physics, Trinity College Dublin, University of Dublin, College Green, Dublin 2, Co. Dublin, D02 PN40, Ireland \\
%$^{2}$ DESY, D-15738 Zeuthen, Germany}

\label{firstpage}

\begin{abstract}
Cosmic rays may have contributed to the start of life on Earth. Here, we investigate the evolution of the Galactic cosmic ray spectrum at Earth from ages $t = 0.6-6.0$\,Gyr. We use a 1D cosmic ray transport model and a 1.5D stellar wind model to derive the evolving wind properties of a solar-type star. At $t=1\,$Gyr, approximately when life is thought to have begun on Earth, we find that the intensity of $\sim$GeV Galactic cosmic rays would have been $\sim10$ times smaller than the present-day value. At lower kinetic energies, Galactic cosmic ray modulation would have been even more severe. 
More generally, we find that the differential intensity of low energy Galactic cosmic rays decreases  at younger ages and is well described by a broken power-law in solar rotation rate. We provide an analytic formula of our Galactic cosmic ray spectra at Earth's orbit for different ages. Our model is also applicable to other solar-type stars with exoplanets orbiting at different radii. Specifically, we use our Galactic cosmic ray spectrum at 20\,au for $t=600\,$Myr to estimate the penetration of cosmic rays in the atmosphere of HR\,2562b, a directly imaged exoplanet orbiting a young solar-type star. We find that the majority of particles $<0.1$GeV are attenuated at pressures $\gtrsim10^{-5}$\,bar and thus do not reach altitudes below $\sim100$\,km. Observationally constraining the Galactic cosmic ray spectrum in the atmosphere of a warm Jupiter would in turn help constrain the flux of cosmic rays reaching young Earth-like exoplanets.
\end{abstract}


\section{Introduction}
\label{sec:intro}

Galactic cosmic rays have been considered as a source of ionisation for exoplanetary atmospheres \citep{rimmer_2013}. Depending on the orbital distance of an exoplanet from its host star it may be possible to disentangle the chemical signature of Galactic cosmic rays from other sources such as stellar radiation and stellar energetic particles. Ionisation by energetic particles, including both Galactic and stellar cosmic rays, is of great interest not only for the chemistry in exoplanetary atmospheres but also at even earlier stages when the protoplanetary disc is still present \citep{cleeves-2013,cleeves-2015,rab_2017,rodgers-lee_2017,rodgers-lee_2020} and for star formation in general \citep[see][for a recent review]{padovani_2020}. 

In terms of the solar system, it is of interest to determine the intensity of Galactic cosmic rays incident on Earth at the time when life is thought to have begun \citep{mojzsis_1996}. Galactic cosmic rays influence and contribute to atmospheric electrical circuits \citep[][in the case of the Earth]{rycroft_2012}, cloud cover \citep{svensmark_2017} and biological mutation rates \citep[see discussion in][for instance]{griessmeier_2005}. Here, we focus on the interaction of Galactic cosmic rays with the stellar winds from {\it solar}-type stars specifically, and note that the effect of Galactic cosmic rays on close-in super-Earth exoplanets around M dwarf stars has also been considered \citep{griessmeier_2005, griessmeier_2009,griessmeier_2015}. We also investigate the Galactic cosmic ray spectrum impinging on exoplanets orbiting young solar-type stars at different orbital distances than the Earth.

The properties of the Sun and its stellar wind are thought to have varied over the lifetime of the Sun. This evolution is inferred from observations of other solar-type stars of different ages since their evolution is thought to be similar. Young solar-type stars typically display much stronger magnetic fields \citep{vidotto_2014, folsom_2016, rosen_2016} and higher X-ray luminosities \citep{wright_2011, tu_2015}, as well as faster rotation rates \citep{gallet_2013}, which are thought to result in higher mass-loss rates via stellar winds \citep{vidotto_2017, ofionnagain_2019}. Thus, since the properties of the solar wind change with time this means that the interaction of Galactic cosmic rays with the solar wind will also vary with time. In this paper we investigate how the solar modulation of Galactic cosmic rays varies as a function of the Sun's life from $0.6-6.0$\,Gyr. This evolution of Galactic cosmic ray modulation should also 

{\it Voyager 1} and {\it 2} measurements have provided us with valuable information about the local interstellar spectrum (LIS) of Galactic cosmic rays outside of the heliosphere \citep{stone_2013,cummings_2016,stone_2019} which are thought to be unaffected by the solar wind. How Galactic cosmic rays then propagate through the magnetised solar wind can be characterised, to first order, as a competitive process between the spatial diffusion of Galactic cosmic rays into the solar system, spatial advection of Galactic cosmic rays out of the system and adiabatic losses of Galactic cosmic rays as they do work against the solar wind \citep{parker_1965}. The suppression of the LIS of Galactic cosmic rays as they travel through the solar wind to Earth is known as the modulation of Galactic cosmic rays. The present-day solar modulation of Galactic cosmic rays that arrive at Earth has been extensively studied \citep{parker_1965,jokipii_1971,potgieter_2013,vos_2015}.

Given that the solar wind has evolved during its main-sequence lifetime, the flux of Galactic cosmic rays arriving at Earth is expected to have changed throughout the Sun's life \citep{svensmark_2006,cohen_2012}. More specifically, \citet{svensmark_2006} used relationships between the solar rotation rate and the magnetic field strength and velocity of the solar wind to estimate these quantities at different times during the Sun's life. \citet{cohen_2012} find that during the Archean eon (approximately the period when life is thought to have started on Earth) that the Earth would have experienced a greatly reduced intensity of Galactic cosmic rays.

Our approach is similar to \citet{svensmark_2006} which uses a 1D transport equation for the Galactic cosmic rays. We build upon this work by using updated observationally derived relationships between the solar rotation rate and the magnetic field strength and velocity of the solar wind. We focus on a number of radii which are relevant for specific exoplanetary systems around solar-type stars. We discuss the differences in results that we find in Section\,\ref{sec:discussion}. 

 In this paper we also focus on the conditions in the early solar system to determine the effect that the Sun being a slow/fast rotator would have. In addition, we estimate the flux of Galactic cosmic rays as a function of radius, focusing on radii of particular interest where the signatures of Galactic cosmic rays in an exoplanetary atmosphere may dominate over other sources of ionisation from a solar-type star (i.e. photoionisation and stellar energetic particles).

Note, the results presented in Section\,\ref{sec:results} mainly discuss the evolution of the GCR spectrum at Earth due to the evolution of the solar wind over the Sun's life. However, the evolution of the GCR spectrum should be similar for other solar-type stars. Thus, in Section\,\ref{subsec:hr2562} we focus on a young solar-type star with a warm Jupiter exoplanet, HR 2562b \citep{konopacky_2016}, orbiting at 20\,au. Assuming an unmagnetised exoplanet, we calculate the energy losses of the cosmic rays as they propagate through the upper atmosphere of the exoplanet.

Finally, we consider the exoplanetary system HR 2562b \citep{konopacky_2016}, assuming an unmagnetised exoplanet, and calculate the energy losses of the cosmic rays as they propagate through the upper atmosphere of the exoplanet. 

The paper is structured as follows: in Section\,\ref{sec:form} we describe the stellar wind model and cosmic ray transport model that we use. We present our results in Sections\,\ref{sec:results} and \ref{subsec:hr2562}. We discuss our results in comparison to other results in the literature in Section\,\ref{sec:discussion}. Finally, we present our conclusions in Section\,\ref{sec:conclusions}.

\section{Formulation}
\label{sec:form}
To model the propagation of Galactic cosmic rays from the interstellar medium (ISM) into the solar system (or into a solar-type star system) we solve the 1D transport equation for the cosmic rays, assuming spherical symmetry, given by
\begin{equation}
\frac{\partial f}{\partial t} = \bm{\nabla}\cdot(\kappa\bm{\nabla} f)-\nabla \cdot(vf) +\frac{1}{3}(\nabla\cdot v)\frac{\partial f}{\partial \mathrm{ln}p} 
\label{eq:f}
\end{equation}
\noindent where $f(r,p,t)$ is the cosmic ray phase space density, $\kappa(r,p)$ is the spatial diffusion coefficient, $v(r)$ is the radial velocity of the stellar wind and $p$ is the momentum of the particles which are taken to be protons. The first term on the righthand side of Eq.\,\ref{eq:f} represents the spatial diffusion of cosmic rays through the stellar wind which depends on the level of turbulence and strength of the magnetic field (described in more detail in Sections\,\ref{subsec:dif} and \ref{subsec:parameters_time}). The second term represents spatial advection which acts to suppress the flux of cosmic rays as they travel into the stellar system. The last term represents momentum advection which pushes the cosmic rays to lower energies as they do work against the magnetised stellar wind to enter the stellar system.

\begin{figure}
	\centering
        \includegraphics[width=0.5\textwidth]{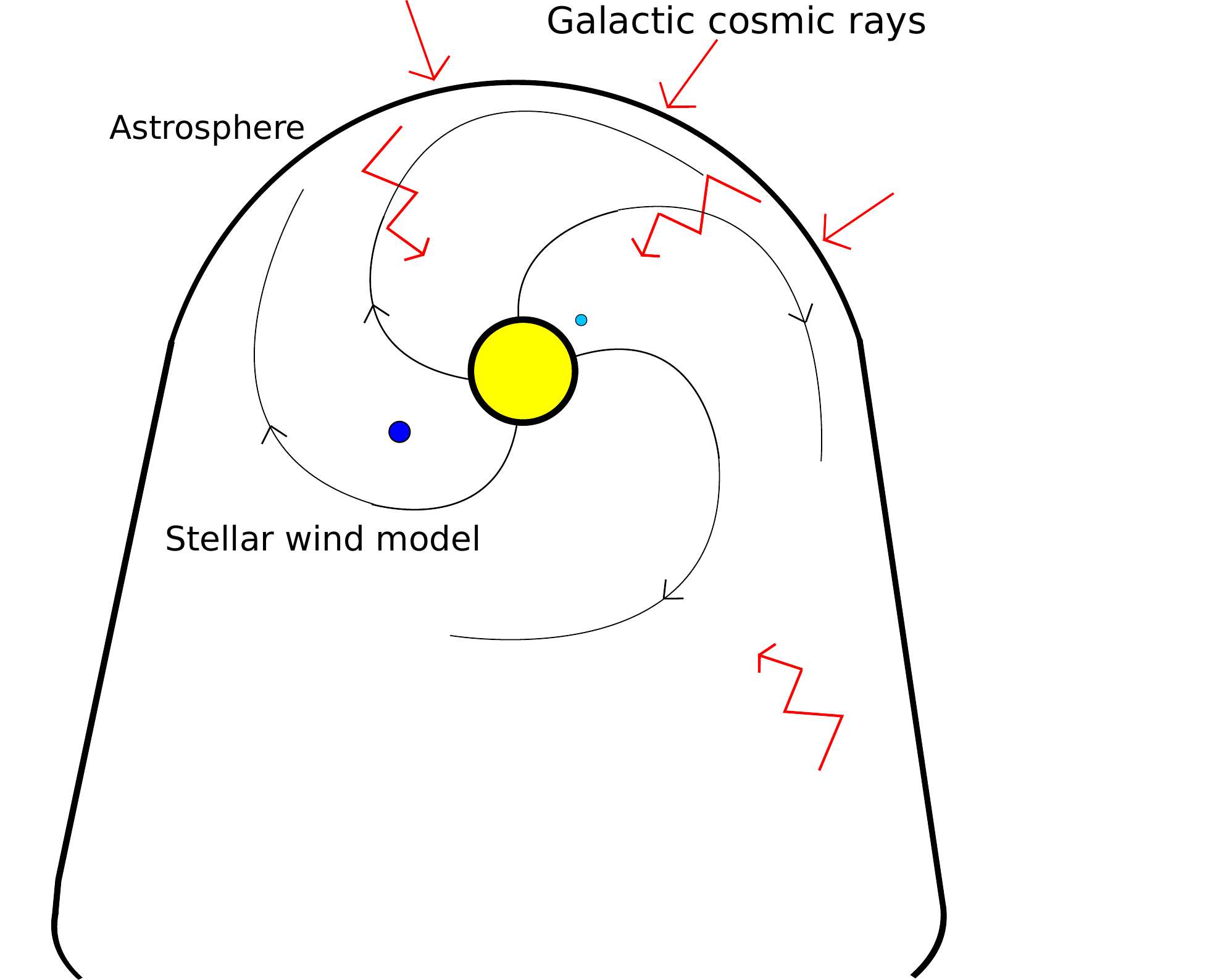}
       	\centering
  \caption{Here we show a schematic of a solar-type stellar system to illustrate Galactic cosmic rays diffusing into a stellar system from outside the astrosphere and eventually arriving at the location of planets. The axisymmetric shape of the astrosphere due to the motion of the star through the ISM is not incorporated in our stellar wind or Galactic cosmic ray model.} 
    \label{fig:stellar_wind}
\end{figure}

Fig.\,\ref{fig:stellar_wind} shows a schematic of the Galactic cosmic rays diffusing into a stellar system from outside the astrosphere. The velocity profile of the stellar wind is derived from the stellar wind model described in Section\,\ref{subsec:parameters_time}. We focus on the steady-state solution of Eq.\,\ref{eq:f} which is a reasonable approximation for solar minimum conditions. The fact that we study the steady-state solution of Eq.\,\ref{eq:f} and also assume azimuthal symmetry means that any short-term modulation effects, shorter than the rotation period of the star, are neglected \citep[see discussion in][]{potgieter_2013}. We also do not include any drift motions of the cosmic rays \citep{jokipii_1977} in Eq.\,\ref{eq:f}. This implies that the known temporal variation of Galactic cosmic ray modulation due to the solar cycle cannot be studied here. The drift motion of the cosmic rays also results in latitudinal variations which we do not consider here. Thus, in the future a more complete study of these effects could be studied using a 2D cosmic ray transport code. These effects should be kept in mind when examining our results and that we are implicitly always investigating solar (or stellar) minimum conditions for these stars. It is also important to note that we also do not consider in our model the effect of the termination shock in the stellar wind and the stellar equivalent of the heliosheath. For the solar system, at $10-100$\,MeV energies $\gtrsim 80\%$ of the modulation of Galactic cosmic rays occurs in the heliosheath \citep[see][for instance]{potgieter_2013}. At the same time, the termination shock can reaccelerate GeV Galactic cosmic rays depending on the magnetic polarity cycle of the Sun. To ascertain how the size and structure of the heliosheath evolves with stellar rotation rate 3D magnetohydrodynamic simulations would be required.

Eq.\,\ref{eq:f} is numerically advanced using a forward in time, second order centred in space differencing scheme for the diffusion term and a first order in space upwinding scheme for the advection terms. The numerical scheme used is overall first order in time. The code that we use is an adapted version of the code presented in \citet{rodgers-lee_2017,rodgers-lee_2020} which now includes momentum advection and a different scheme for the advection terms. A full description of the numerical scheme is given in Appendix\,\ref{appendix:numerics} including the implementation of the boundary conditions which is described in Appendix\,\ref{appendix:bcs}. We validate our code by showing that it reproduces well observations of Galactic cosmic rays measured at Earth which is presented in Appendix\,\ref{appendix:observations}. A numerical convergence test for the scheme is given in Appendix\,\ref{appendix:resolution}. 

For the boundary conditions, the spatial inner boundary condition is reflective. We use a fixed spatial outer boundary condition with the boundary cell taken to be the LIS value, described in Section\,\ref{subsec:lis}. The momentum inner and outer boundary conditions are outflow.

\begin{table*}
\centering
\caption{List of parameters for the simulations. The columns are, respectively: the age ($t$) of the Sun, its rotation rate ($\Omega$) in terms of the present-day value ($\Omega_\odot = 2.67\times 10^{-6}\,\mathrm{rad\,s^{-1}}$), its rotation period ($P_\mathrm{rot}$), the heliospheric radius ($R_\mathrm{h}$, Eq.\,\ref{eq:rh}), the radial velocity ($v_{1\mathrm{au}}$) and the magnitude of the total magnetic field ($|B_{1\mathrm{au}}|$) at $r=1$\,au. $|B_*|$ and $T_*$ are the magnitude of the total magnetic field and the temperature at the base of the wind ($r_\odot$) and $\dot M$ is the mass-loss rate. The final column is the potential ($\phi$, from the modified force field approximation, Eq.\,\ref{eq:force-field}), that we find for each of the simulations. 
}
\begin{tabular}{@{}cccccccccc@{}}
\hline

$t$ &$\Omega$ &$P_\mathrm{rot}$& $R_\mathrm{h}$ & $v_{1\mathrm{au}}$ &$|B_{1\mathrm{au}}|$&$|B_*|$ & $T_*$&  $\dot M$ &  $\phi$\\
\hline
[Gyr] & $[\Omega_\odot]$&[days]   &[au] & $\mathrm{[km\,s^{-1}]}$ & [G] & [G] & [MK] &[$M_\odot\,\mathrm{yr}^{-1}$] & [GeV]\\
\hline
6.0 & 0.87	 & 31   & 47    & 370 &$3.4\times 10^{-5}$ &1.1 & 1.3 & $4.1\times 10^{-15}$ & 0.09\\ 
4.6 			 & 1.0 	 & 27   & 122  & 450 &$3.8\times 10^{-5}$ &1.3 & 1.5 & $2.3\times 10^{-14}$ & 0.21\\ 
2.9 			 & 1.3 	 & 22   & 500  & 610 &$5.4\times 10^{-5}$ &1.7 & 2.2 & $2.8\times 10^{-13}$ & 0.57 \\ 
1.7 			 & 1.6 	 & 17   & 696  & 660 &$7.6\times 10^{-5}$ &2.5 & 2.4 & $5.1\times 10^{-13}$ & 1.19\\
1.0 			 & 2.1   & 13   & 950  & 720 &$1.2\times 10^{-4}$ & 3.5 & 2.6 & $8.5\times 10^{-13}$ &1.96\\
\hline
0.6 			 & 3.0	 & 9    & 1324 & 790 &$1.8\times 10^{-4}$ &5.5 & 3.0 & $1.5\times 10^{-12}$ & 5.1$^\dag$\\ 
0.6 			 & 3.5   & 8    & 1530 & 820 &$2.8\times 10^{-4}$ &6.7 & 3.2 & $2.0\times 10^{-12}$ & 7.45$^\dag$\\ 
0.6 			 & 4.0 	 & 7    & 1725 & 850 &$3.5\times 10^{-4}$ &8.0 & 3.3 & $2.4\times 10^{-12}$ &10.3$^\dag$\\
\hline \\
\end{tabular}

\dag These values for $\phi$ do not match our results very well below the peak of the Galactic cosmic ray spectrum (Section\,\ref{subsubsec:force-field}). 
\label{table:sim_parameters}
\end{table*}

\subsection{Local interstellar spectrum (LIS)}
\label{subsec:lis}
The LIS of Galactic cosmic rays is the spectrum that is thought to be unmodulated by the solar wind and therefore can only be observed outside of the heliosphere. The LIS has been measured by \emph{Voyager 1} from beyond the heliopause \citep{stone_2013,cummings_2016}. A model fit to the \emph{Voyager 1} observations of the LIS, from \citet{vos_2015}, is given as a differential intensity, $j_\mathrm{LIS}$, as
\begin{equation}
j_\mathrm{LIS}(T) = 2.70\frac{T^{1.12}}{\beta^2}\left(\frac{T+0.67}{1.67} \right)^{-3.93} \mathrm{m^{-2}}\,\mathrm{s^{-1}}\,\mathrm{sr^{-1}}\,\mathrm{MeV^{-1}}
\label{eq:lis}
\end{equation}
where $T$ is the kinetic energy of the cosmic rays in GeV and $\beta$ is the velocity of the particle divided by the speed of light $c$. In the model of \citet{vos_2015} the very LIS is specified at the heliopause, taken to be 122\,au. For our simulations the value at the outer boundary is taken to be the LIS\footnote{The expression for the LIS in Eq.\,\ref{eq:lis} is different at low energies from the LIS used in \citet{svensmark_2006} and \citet{cohen_2012} which is based on a model fit to older \emph{Voyager 1} data. At $\sim1$\,GeV and higher energies the spectra are the same but below $\sim1$\,GeV the model fit from \citet{vos_2015} is now more accurate as it is constrained by the more recent \emph{Voyager 1} data. However, since the difference in the adopted spectra is only at low energies where solar modulation dominates it is unlikely that the different spectra would affect the model results.}, where the differential intensity of cosmic rays can be expressed in terms of the phase space density ($f$ from Eq.\,\ref{eq:f}) as $j(T) = p^2f(p)$.

We assume a constant LIS as a function of time in our simulations. The LIS may have evolved as a function of time, due to a corresponding temporal evolution of the star formation rate (SFR) of the Milky Way \citep{rocha-pinto_2000} and assuming that the majority of Galactic cosmic rays are produced by supernovae, as discussed in \citet{svensmark_2006}. However since the Milky Way's SFR for the times that we consider ($t>0.6\,$Gyr), shown in Fig.\,2 of \citet{svensmark_2006}, is within a factor of two of the current value for the Milky Way's SFR, we do not vary the LIS as a function of time.

\subsection{Diffusion coefficient}
\label{subsec:dif}
The diffusion coefficient of the cosmic rays, in units of $c$, can be estimated from quasi-linear theory \citep{jokipii-1966,schlickeiser_1989} as 
\begin{equation}
\frac{\kappa(r,p,\Omega)}{\beta c} =\eta_0 \left( \frac{p}{p_0}\right)^{1-\gamma}r_\mathrm{L}
\label{eq:kappa}
\end{equation}
where $r_\mathrm{L} =p/[eB(r,\Omega)]$ is the Larmor radius of the protons with $e$ representing the unit of electric charge, $\Omega$ is the adopted stellar rotation rate and
\begin{equation}
\eta_0 = \left(\frac{B}{\delta B} \right)^2
\label{eq:eta0}
\end{equation}
where $B^{2}$ relates to the energy density of the large-scale magnetic field and $(\delta B)^{2}$ to the total energy density in the smaller scale magnetic field turbulent modes. The diffusion coefficient $\kappa/\beta c$ describes the scattering length of protons with momentum $p$, and in Eq.\,\ref{eq:kappa} is scaled to momentum $p_{0}$, corresponding to momentum of particles whose Larmor radii matches the length of the longest turbulent modes. We adopt $p_{0}=3\,\mathrm{GeV}/c$. The value of $\eta_0$ represents the level of turbulence present in the magnetic field (Eq.\,\ref{eq:eta0}). The value of $\gamma$ is related to the turbulence power spectrum where $\gamma=5/3$ would represent Kolmogorov-type turbulence. The value of $\gamma=1$ was adopted by \citet{svensmark_2006} and \citet{cohen_2012} which fits the present day observations of solar wind modulation quite well and which we also show in Fig.\,\ref{fig:standard} using $\eta_0=1$. Thus, we adopt $\eta_0=1$ and $\gamma=1$ for all of the simulations. The magnetic field strength of a solar-type star increases with increasing stellar rotation rate (which is discussed further in Section\,\ref{subsubsec:vel_mag}). Given that we adopt a constant value for $\eta_0$, this means that the diffusion coefficient decreases with increasing magnetic field strength and therefore also with increasing stellar rotation rate. The possible implications of these assumptions are discussed briefly in Appendix\,\ref{subsec:eta0_gamma}. The solar wind properties that we adopt for the present day simulation ($t=4.6\,$Gyr) are given in Table\,\ref{table:sim_parameters} and are also described in the subsequent sections.

\subsection{Stellar wind parameters as a function of time}
\label{subsec:parameters_time}
A number of physical quantities relating to the wind of a solar-type star must be defined in order to solve Eq.\,\ref{eq:f}, namely the velocity and magnetic field profile as a function of radius and time, as well as the heliospheric radius. Here, we describe our stellar wind model to simulate the long-term evolution of the  wind of a solar-type star, based on empirical relations derived from samples of solar-type stars. In our model, we use rotation as a proxy for age, so that young solar-type stars rotate faster than more evolved solar-type stars. The term ``solar-type'' star is often used to refer to low-mass stars with masses in the range of $\sim0.5 - 1.3\,M_\odot$ corresponding to low-mass stars with convective envelopes. We run our stellar wind model only for stars with $M_* = 1\,M_\odot$ to be able to focus on the Sun's evolution. Thus, it can also be applied to stars with similar masses. 

The stellar wind model that we use to derive the stellar wind properties as a function of radius for different ages is a 1.5D Weber-Davis model \citep{weber_1967}, which assumes that the star is rotating and magnetised. The code that we use which implements this magneto-rotator model is presented in \citet{johnstone_2015a} and \citet{carolan_2019}, based on the Versatile Advection Code \citep[VAC,][]{toth_1996}. We assume that the magnetic field, temperature and density at the base of the stellar wind scale with the stellar rotation rate \citep{carolan_2019}. The surface of the Sun is located at $1\,r_\odot$ (i.e. one solar radius) corresponding to the photosphere while the corona is located at $\sim1.003\,r_\odot$, slightly above the photosphere. Our stellar wind model launches the wind from the base of the corona which we approximate as $1\,r_\odot$. For any given rotation rate, the stellar wind model then solves for the distance profiles of the magnetic field (the radial and azimuthal components), radial and azimuthal velocity, pressure and mass density. The resulting radial profiles for the relevant physical quantities are then used in Eq.\,\ref{eq:f}.

Our stellar wind model is polytropic meaning that the pressure is related to the density via $P \propto \rho^\alpha$, where we assume here that $\alpha=1.05$. Therefore, the stellar wind temperature profile is close to being isothermal. The polytropic wind model assumes that the driving mechanism for the solar wind is thermal pressure gradients. More details of our adopted stellar wind model are shown in \citet{carolan_2019}. It is important to note that the physical properties that we derive from the stellar wind model are applicable to the Sun and also to other solar-type stars. Therefore, throughout the paper we often refer more generally to stellar winds rather than to the solar wind since our results are equally valid for other solar-type stars.

\subsubsection{Stellar rotation rate as a proxy for age}
The evolution in time of the rotation rate for a solar-type star can be derived from large observational samples of solar-type stars with different ages \citep[Fig.\,3,][]{gallet_2013}. At very young ages ($t<5-10\,$Myr), the presence of protoplanetary discs brake the spin up of young stars that would otherwise occur due to gravitational contraction. Once protoplanetary discs are dispersed young stars then continue to spin up at a faster rate until they reach the zero-age main sequence. After this, the spin down of solar-type stars is attributed to stellar winds, which carry away angular momentum. We limit our study to ages $t \geq 0.6\,$Gyr, as some of our assumptions for the properties of the stellar wind base may no longer hold at very young ages.

From $\sim 0.6$ to $1$\,Gyr, observations show a large spread in rotation rates of solar-type stars, which means that prior to $\sim 1$\,Gyr it is not possible to determine the rotation rate of the Sun (e.g. fast or slow rotator). Therefore, for $t=0.6$\,Gyr we investigate three scenarios, ranging from the case where the Sun was a slow rotator ($\Omega = 3\Omega_\odot$) to a fast rotator ($\Omega = 4\Omega_\odot$) scenario, with an intermediate rotator case of $\Omega = 3.5\Omega_\odot$. However, after $\sim 1$\,Gyr (corresponding to $\sim2.1\,\Omega_\odot$), the rotation rate of the Sun is thought to have converged, such that  $\Omega \propto t^{-0.5}$ \citep{skumanich_1972}. 

The values of $\Omega$ that we investigate here, as well as the age and other corresponding physical parameters of our simulations, are given in Table\,\ref{table:sim_parameters}. We simulate the evolving solar wind for the following rotation rates: $\Omega = 0.87,\,1.0,\,1.3,\,1.6,\,2.1,\,3.0,\,3.5,\,4.0\,\Omega_\odot $.

\subsubsection{The evolving winds of solar-type stars}
\label{subsubsec:vel_mag}
Magnetic torques in the winds of solar-type stars are responsible for carrying away most of the stellar angular momentum. To prescribe the evolution of the magnetic field for solar-type stars, we use the empirical relationship between observationally derived values of the large-scale magnetic field strength for low-mass stars and stellar rotation rate \citep{vidotto_2014} given by
\begin{equation}
B_{*}(\Omega) = 1.3 \left(\frac{\Omega}{\Omega_\odot}\right)^{1.32\pm0.14}\,\mathrm{G}.
\label{eq:b}
\end{equation}
The field strength was obtained by averaging surface magnetic maps, which for stars was derived using the Zeeman Doppler Imaging (ZDI) technique. For the Sun the large scale component of solar synoptic maps (derived from Kitt Peak/National Solar Observatory data) was instead used. 

We use these observationally derived values for $B_{*}(\Omega)$ as the value of the radial component of the magnetic field strength, $B_{*,r}(\Omega)$, at the wind base for the stellar wind model. The initial condition used in the stellar wind model for the radial profile of the magnetic field is that $B_r(r,\Omega) = B_{*,r}(\Omega)(r_\odot/r)^2 $ and $B_\phi(\Omega) = 0$. As the stellar wind simulation evolves, an azimuthal component of the magnetic field  develops due to stellar rotation. At large distances, this component falls off as $1/r$. Our steady-state stellar wind models extend out to 1\,au. \citet{carolan_2019} showed that the magnetic field at Earth's orbit from this stellar wind model ($\sim3.8\times10^{-5}$G) matches the observed values very well. \citet{finley_2019}, for example, shows the observed open magnetic flux in the solar wind varying from $5-15\times 10^{22}$Mx, which results in magnetic field strengths at Earth's orbit of $1.78-5.3\times10^{-5}$G. The model also matches the observed values for the mass-loss rate, velocity and density of the solar wind at 1\,au very well, as shown in Fig.1 of \citet{carolan_2019}.

We extrapolate the values of $B_r(\Omega)$ and $B_\phi(\Omega)$ beyond $1\mathrm{au}$ out to the edge of the heliosphere using power laws with distance such that 
\begin{eqnarray}
B_r(r>1\mathrm{au},\Omega) &=& B_{r,1\mathrm{au}}(\Omega)\left(\frac{1\mathrm{au}}{r}\right)^2 \\
B_\phi(r>1\mathrm{au},\Omega) &=& B_{\phi,1\mathrm{au}}(\Omega)\left(\frac{1\mathrm{au}}{r} \right)
\end{eqnarray}
\noindent Since the radial magnetic field falls as $1/r^2$ but the azimuthal field only decreases as $1/r$, this gives rise to the Parker spiral that becomes tighter at larger distances when $B_\phi$ dominates. The values we obtain for the total magnetic field strength as a function of orbital distance and stellar rotation rate are used to determine the diffusion coefficient given in Eq.\,\ref{eq:kappa}. A fit to the values of $B_{\phi,1\,\mathrm{au}}$ and $B_{r,1\,\mathrm{au}}$ as a function of stellar rotation rate, derived from the stellar wind model values, is given in Eq.\,A1 of \citet{carolan_2019} in combination with the values quoted in their Tables\,A1-A2.

Note, we use the best fit values for the magnetic field strength as a function of stellar rotation rate given in Eq.\,\ref{eq:b} and thus we do not consider the effect of the uncertainty in the fit here. Note that ZDI only allows the large-scale field to be reliably reconstructed \citep{johnstone_2010, arzoumanian_2011, lang_2014}. Fortunately, the stellar wind flows through large-scale fields and therefore the limited resolution of ZDI magnetograms has been demonstrated not to affect the stellar wind \citep{jardine_2017, boro-saikia_2020}. \citet{lehmann_2019} performed a study of the ZDI technique using controlled input data and showed that the large-scale field morphologies are recovered well.

The other two wind base parameters required in our stellar wind models are the base temperature and density. We  use the relationship for the stellar wind base temperature as a function of rotation rate from \citet{ofionnagain_2018}:
\begin{equation}
T_* = \begin{cases} 1.50 \left( \frac{\Omega}{\Omega_\odot}\right)^{1.2} \mathrm{MK} \hspace{4mm}\mathrm{for} \hspace{2mm} \Omega < 1.4\Omega_\odot \\
					 1.98\left( \frac{\Omega}{\Omega_\odot}\right)^{0.37} \mathrm{MK\,}\hspace{2mm} \mathrm{\,for} \hspace{3mm} \Omega \geq 1.4\Omega_\odot
\end{cases}
\end{equation}
For the base density, we assume that $n_* = 10^{8}  (\Omega/\Omega_\odot)^{0.6} $cm$^{-3}$, following the work by \citet{ivanova_2003}. 

Overall, the radial velocity profile results from the magneto-rotator stellar wind model that we use given a particular set of values for the temperature, density and magnetic field strength at the base of the wind. The stellar wind model, by construction, matches the solar wind velocities observed at Earth well \citep[$v_\oplus \simeq 450\,\mathrm{km\,s^{-1}}$,][]{mccomas_2008,usmanov_2014}. For each of the stellar wind simulations the wind has reached its terminal velocity by 1\,au and so $v(r>1\mathrm{au}) =v(1\mathrm{au}) $ is used in Eq.\,\ref{eq:f}. The values of $v$ at $1\mathrm{au}$ (denoted by $v_{1\mathrm{au}}$) are given in Table\,\ref{table:sim_parameters}. Fig.\,\ref{fig:profiles} in Appendix\,\ref{appendix:profiles} shows the magnetic field strength and velocity profiles, as a function of radius, derived from the magneto-rotator stellar wind model for $1\,\Omega_\odot$ and $4\,\Omega_\odot$.

From mass conservation, it follows that $\dot{M} = 4\pi r^2 m n v$, with $m=0.5 m_\mathrm{p}$ being the mean mass of the solar wind particle, considered to be composed of fully ionised hydrogen. At the present-day solar rotation rate, our model assumptions reproduce the present-day value of the solar wind mass-loss rate: $\dot{M} = 2\times 10^{-14} M_\odot\,\mathrm{yr^{-1}}$.  The mass-loss rates calculated at other ages are shown in Table\,\ref{table:sim_parameters} and are used to calculate the heliospheric (or more generally the astrospheric) radius. 

\subsubsection{Heliospheric radius}
\label{subsubsec:rh}
The radius of a solar-type star's astrosphere, $R_\mathrm{h}$, is determined as a balance of the stellar wind ram pressure ($P_\odot\propto\dot M_\odot v_\odot/r^2$) and the ambient ISM pressure, $P_\mathrm{ISM}$. The solar wind ram pressure evolves with time and so, by assuming a constant ISM pressure as a function of time \citep[following][]{svensmark_2006}, we can estimate $R_\mathrm{h}$ as a function of time as
\begin{equation}
R_\mathrm{h}(t) = R_\mathrm{h,\odot} \sqrt{\frac{{\dot M(t)}v(t)}{{\dot M_\odot}v_\odot}}
\label{eq:rh}
\end{equation}
\noindent where $R_\mathrm{h,\odot}$, $\dot M_\odot$ and $v_\odot$ are the values for the Sun's current heliospheric radius, mass loss rate and wind velocity at $1\mathrm{au}$, respectively. The present day values for these parameters are given in Table\,\ref{table:sim_parameters}, as well as the values for different times. 

\subsection{Our combined stellar wind and cosmic ray propagation simulations}
We use the output of our stellar wind simulations in our simulations of cosmic ray propagation. To recapitulate,
we run a number of stellar wind simulations for a number of different times during a solar-type star's life (Table\,\ref{table:sim_parameters}). For each time, we obtain the stellar wind velocity and the magnetic field profile from the wind base at $r_\odot$ out to 1\,au, as well as the corresponding mass-loss rate. Beyond 1\,au, we use the fact that the stellar wind has reached terminal speed and extrapolate the stellar wind conditions out to the astrospheric radius. 

For all the cosmic ray propagation simulations, the inner radial spatial boundary is set to $0.1\,$au. We use the mass-loss rate and the radial velocity of the stellar wind in Eq.\,\ref{eq:rh} to derive the heliospheric radius as a function of time. Thus, our outer radial boundary is set to the heliospheric radius, $R_\mathrm{h}(t)$. Therefore, the logarithmically spaced radial bins for $i=0,...,N$ are given by $r_i = \mathrm{exp}\{i\times\mathrm{ln}(r_N/r_0)/(N-1) + \mathrm{ln}\,r_0\}$ where $r_0=0.1\,$au and $r_N=R_\mathrm{h}(t)$ with $N=60$. Similarly, $p_j = \mathrm{exp}\{j\times\mathrm{ln}(p_M/p_0)/(M-1) + \mathrm{ln}\,p_0\}$  for $j=0,...,M$ represent the logarithmically spaced momentum bins for the cosmic rays with $M=60$. The minimum and maximum momenta of the cosmic rays that we consider are $p_0=0.15\,\mathrm{GeV}/c $ and $p_M=100\,\mathrm{GeV}/c$, respectively. The same range in momentum is used for all of the cosmic ray propagation simulations.

\section{Results}
\label{sec:results}

We present the results of our numerical study which investigates how the modulation of Galactic cosmic rays by the wind of a solar-type star would evolve throughout a solar-type star's lifetime. We investigate the evolution of the cosmic ray intensity for a number of different cosmic ray energies with the stellar rotation rate. We then specifically look at the radial dependence of the Galactic cosmic ray spectrum at $\sim1\,$Gyr when life is thought to have begun on Earth. We also focus on the differential intensity of Galactic cosmic rays at $t=600\,$Myr which is relevant for the warm Jupiter exoplanet, HR\,2562b, orbiting a solar-like star at a distance of 20\,au. 

\subsection{Galactic cosmic ray spectrum as a function of time}
\label{subsec:gcr_time}

\begin{figure}
	\centering
        \includegraphics[width=0.5\textwidth]{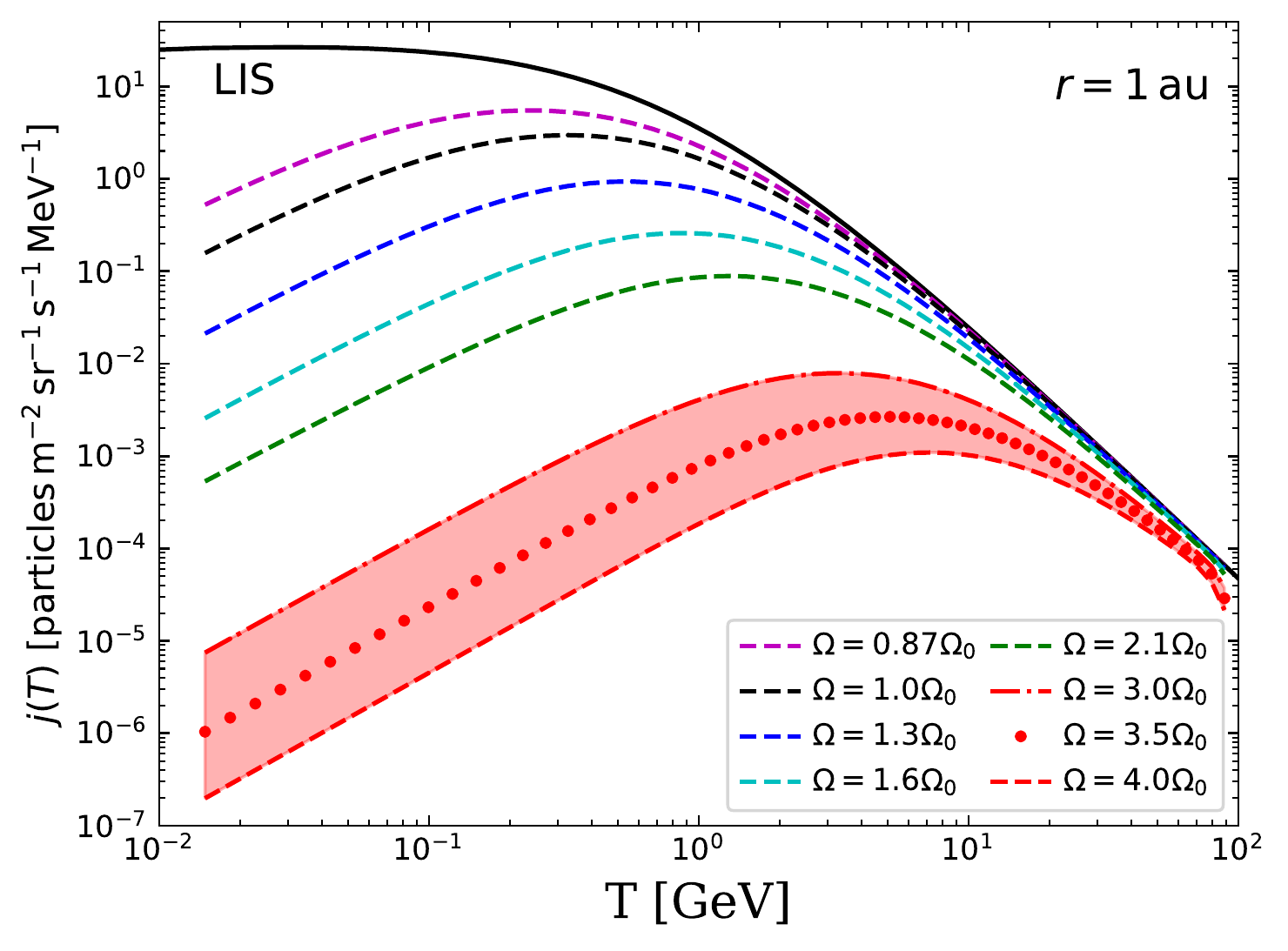}
       	\centering
  \caption{Differential intensity of Galactic cosmic rays as a function of kinetic energy at 1\,au for different stellar rotation rates, which approximately correspond to different ages during the Sun's life. The values of $\Omega$ plotted correspond to $t = 0.6-6.0$\,Gyr for the Sun. The solid black line represents a model fit of the \textit{Voyager 1} data for the LIS (Section \ref{subsec:lis}) which is set to be the value at the outer boundary of the simulations.  The parameters used for each simulation are given in Table\,\ref{table:sim_parameters}.} 
    \label{fig:omega}
\end{figure}

We investigate the Galactic cosmic ray spectrum at the orbital distance of Earth as a function of a solar-type star's lifetime. We focus on a number of different times ranging from $0.6-6.0$\,Gyr which are given in Table\,\ref{table:sim_parameters}. We chose to investigate $t=1.0\,$Gyr as this approximately matches the time at which life is thought to have started on Earth \cite[3.8\,Myr ago,][]{mojzsis_1996}. It is therefore of interest to estimate the intensity of Galactic cosmic rays at this time. The other time of particular interest that we focus on is $t=0.6\,$Gyr since there are observations of a directly imaged exoplanet (HR 2526b) orbiting a star similar in mass to the Sun with an age estimate of $t \sim 0.6\,$Gyr. The impact of Galactic cosmic rays in this exoplanetary system will be discussed further in Section\,\ref{subsec:hr2562}.

Fig.\,\ref{fig:omega} shows the differential intensity of Galactic cosmic rays as a function of their kinetic energy for a number of different stellar rotation rates at 1\,au. The black dashed line represents the present day values that we calculate and the solid black line represents the LIS which is the adopted value of the fixed outer spatial boundary condition.

The magenta dashed line represents a solar-type star with a slower rotation rate ($\Omega = 0.87\Omega_\odot$) than the Sun's present day value and thus probes the intensity of Galactic cosmic rays in the future when the Sun will be $\sim$6.0\,Gyr old. The stellar wind properties present at this time (derived from the stellar wind model) will result in an increase in the number of $\lesssim$GeV cosmic rays reaching Earth, ranging from a factor of $\sim2$ up to a factor $\sim5$ for MeV cosmic rays.

Examining the intensity of Galactic cosmic rays for faster stellar rotation rates, looking into the Sun's past, shows that the intensity decreases rapidly for all but the most energetic cosmic rays. The peak in the differential Galactic cosmic ray intensity as a function of increasing stellar rotation shifts to higher energies as a result of the corresponding increase in the stellar magnetic field strength (which will result in smaller diffusion coefficients) combined with the effect of larger stellar wind velocities.

The red shaded region represents three simulations at $t = 0.6\,$Gyr. Because of the uncertainty of the rotation rate of the Sun at that time, we adopt three values of rotation rate $\Omega = 3.0, 3.5$ and $4\Omega_\odot$. This indicates that at young ages, for $T\lesssim 5$\,GeV, there is at least an order of magnitude difference in the differential intensity of Galactic cosmic rays that reached Earth, depending on whether the Sun was a fast or a slow rotator. Again, it is important to note that we do not include the drift motion of the Galactic cosmic rays in our simulations which, depending on the solar cycle, would lead to a change in our results.
\begin{figure*}%
	\centering
    \subfigure[ ]{%
        \includegraphics[width=0.475\textwidth]{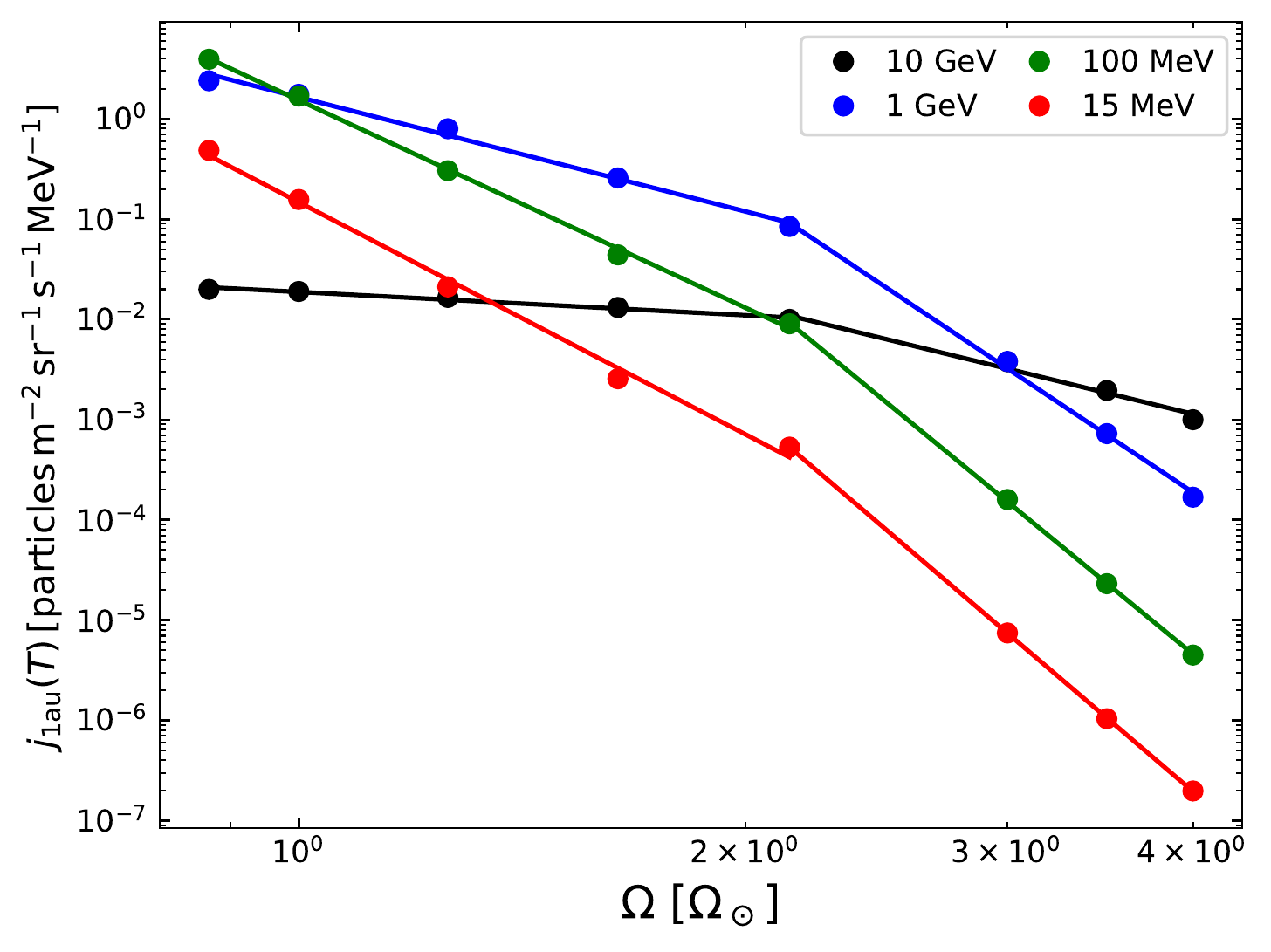}
       	\centering
\label{fig:time_flux}}%
  	    ~
    \subfigure[ ]{%
        \includegraphics[width=0.525\textwidth]{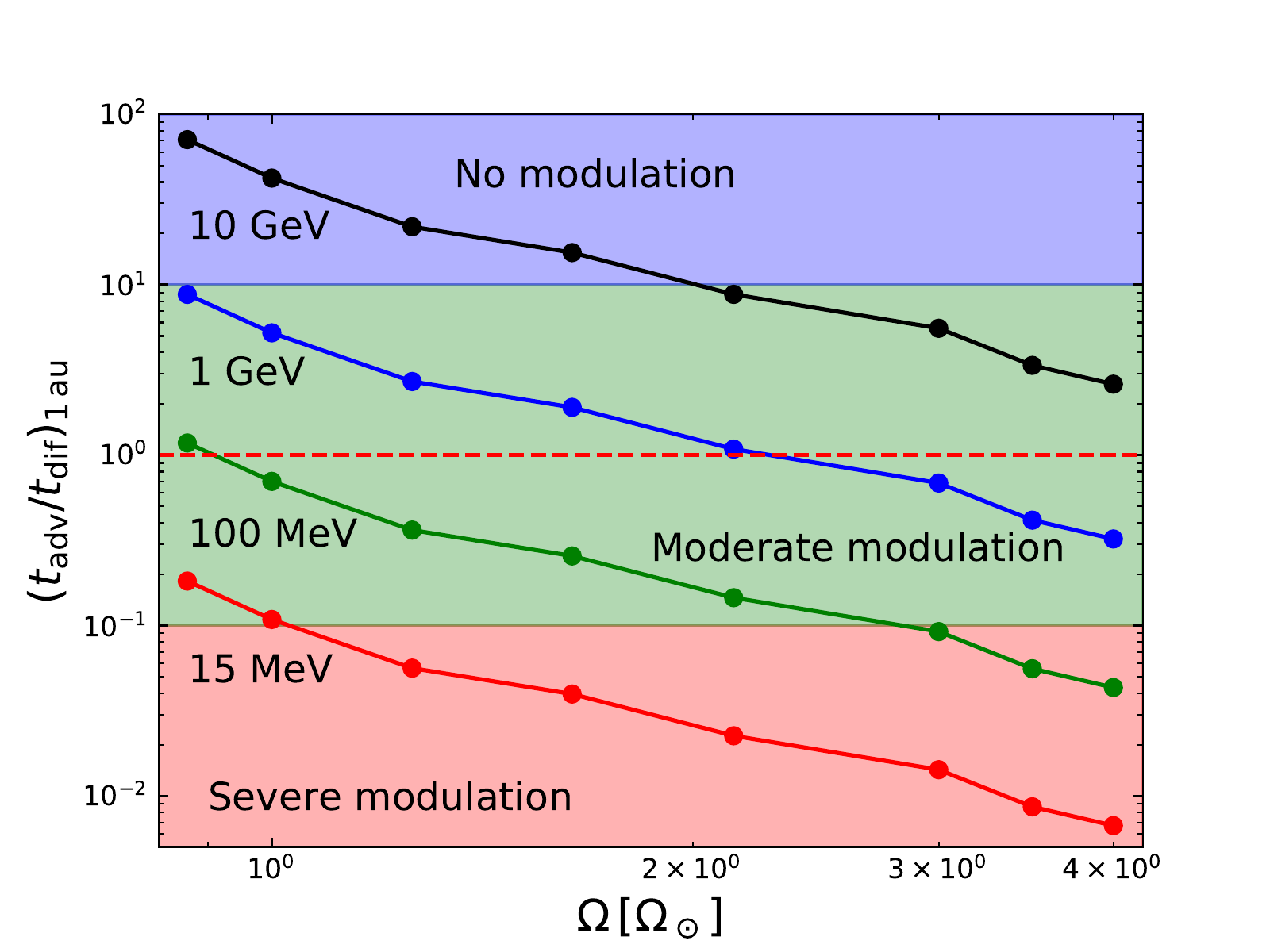}
	\centering
 \label{fig:timescale_compare}}%
    \caption{(a) Differential intensity of Galactic cosmic rays at 1\,au as a function of stellar rotation rate, $\Omega$, for cosmic rays of different kinetic energies. (b) Comparison of the diffusive timescale with the advective timescale as a function of a solar-type star's rotation rate for cosmic rays with different kinetic energies. } 
    \label{fig:flux_timescale}%
\end{figure*}

\subsubsection{Modified force field approximation}
\label{subsubsec:force-field}

The force field approximation \citep{gleeson_1968} provides a simple analytic expression which depends only on a modulation potential, $\phi$, that was developed to describe the solar modulation of Galactic cosmic rays. Here, we compare our results in Fig.\,\ref{fig:omega} with a modified version of the force field approximation because the canonical force field approximation does not fit our simulations well at $\sim$MeV energies\footnote{The fact that the force field approximation does not fit the low energy component of the Galactic cosmic ray spectrum at Earth was first noted by \citet{gleeson_1973} and is also discussed in detail in \citet{caballero-lopez_2004}. }. This modified force field approximation for the differential intensity of Galactic cosmic rays at Earth, $j_\mathrm{1au}(T)$, can be expressed as:
\begin{equation}
\label{eq:force-field}
\frac{j_\mathrm{1au}(T)}{E^2-E_p^2} =  \beta\left(\frac{ j_\mathrm{LIS}(T+\phi)}{(E+\phi)^2-E_p^2} \right)
\end{equation} 
where $E$ is the proton energy and $E_p=0.938$\,GeV is the proton rest energy. The difference between this modified force field approximation and the usual force field approximation is the factor of $\beta$ on the right hand side of Eq.\,\ref{eq:force-field} which increases the suppression at low energies. For the usual force field approximation, $\phi$ is effectively the average energy loss suffered by a cosmic ray reaching Earth coming in from infinity, i.e. the ISM. The values of $\phi$ which fit our data best are given in Table\,\ref{table:sim_parameters}. For $\Omega\gtrsim 2.1 \Omega_\odot$ the modified force field approximation does not fit the low energy cosmic ray intensities very well (see Fig.\,\ref{fig:ffa_comparison} in Appendix\,\ref{appendix:ffa} for a comparison between the modified force field approximation and our results). On the other hand, for $\Omega\lesssim 2.1 \Omega_\odot$ the modified force field approximation, along with the values of $\phi$ quoted in Table\,\ref{table:sim_parameters}, can be used to well approximate our results at 1\,au. It is also important to note that while the (modified) force field approximation can be used to well reproduce the Galactic cosmic ray spectrum at Earth for $\Omega\lesssim 2.1 \Omega_\odot$, it fails to reproduce the Galactic cosmic ray spectrum at large radii \citep[as discussed in][]{caballero-lopez_2004}.

\subsection{Intensity at Earth as a function of time for different energies}
Fig.\,\ref{fig:time_flux} shows the differential intensity of the cosmic rays at 1\,au as a function of $\Omega$, for a number of different kinetic energies. As expected, the lowest energy cosmic rays show the largest decrease in intensity as a function of increasing rotation rate. 

For $T = 0.015-10$\,GeV a similar evolution with increasing rotation rate is observed. Using a least-squares fitting method, we find that the intensity of 1\,GeV cosmic rays decreases as $\Omega^{-3.8}$ until $\Omega\sim 2\Omega_\odot$. For $\Omega \gtrsim 2\Omega_\odot$ the intensity decreases more rapidly following a power law of $\Omega^{-9.9}$. For $T=10\,$GeV, the modulation is relatively small until $\Omega\sim 2\Omega_\odot$ in comparison to the lower energy cosmic rays.  

The break in the power laws at $\Omega\sim 2\Omega_\odot$ can be understood by comparing the diffusive and advective timescales at 1\,au, shown in Fig.\,\ref{fig:timescale_compare}, where
\begin{equation}
t_\mathrm{dif} = \frac{r^2}{\kappa(r,p,\Omega)},\hspace{5mm} t_\mathrm{adv} = \frac{r}{v(r,\Omega)}.
\end{equation}
The diffusion timescale depends on the momentum of the cosmic rays whereas the advective timescale does not. Thus, for any given value of $\Omega$ in Fig.\,\ref{fig:timescale_compare} the variation in the ratio of $t_\mathrm{adv}/t_\mathrm{dif}$ as a function of cosmic ray energy occurs because $\kappa \propto p$. The break in the power law occurs at the same rotation rate for all low-energy cosmic rays. In particular, it occurs approximately when $t_\mathrm{adv}/t_\mathrm{dif} \lesssim 1$ for GeV cosmic rays. The timescales for GeV cosmic rays determines the position of the power law break because cosmic rays with lower energies will always be related to higher energy cosmic rays via momentum advection (i.e. losses). Looking at the LIS spectrum, the differential intensity of $>$GeV cosmic rays is always lower than the intensity of $\lesssim$GeV energies and thus are unable to replace the GeV cosmic rays via momentum advection that are suppressed by spatial advection. 

Fig.\,\ref{fig:timescale_compare} can be used to broadly understand the overall modulation of Galactic cosmic rays. For 10\,GeV cosmic rays because their diffusive timescale is much shorter than the advective timescale they do not experience much modulation until sufficiently far into the Sun's past when the magnetic field strength and the velocity of the solar wind have increased significantly. For GeV and MeV cosmic rays their diffusive timescales (for the present day Sun and the past physical values of the solar wind) are always close to, or longer than, the advective timescale. Thus, the modulation of Galactic cosmic rays with these energies as a function of the Sun's lifetime has always been quite significant. In the future if the magnetic field strength and velocity of the solar wind continue to decrease the differential intensity of Galactic cosmic rays at Earth will converge towards the LIS, with $j(\mathrm{MeV})>j(\mathrm{GeV})$. The magenta dashed line in Fig.\,\ref{fig:omega} shows the differential intensity of cosmic rays at Earth in the future for $t=6.0\,$Gyr, which is still at this time strongly suppressed by the solar wind at low energies. 

Note that the momentum advection term in Eq.\,\ref{eq:f} also has an associated timescale but it will always be longer than the spatial advection timescale and therefore would not be responsible for the observed power law break.

\subsection{Galactic cosmic ray spectrum at the time when life is believed to have started on Earth}
\begin{figure}
	\centering
        \includegraphics[width=0.5\textwidth]{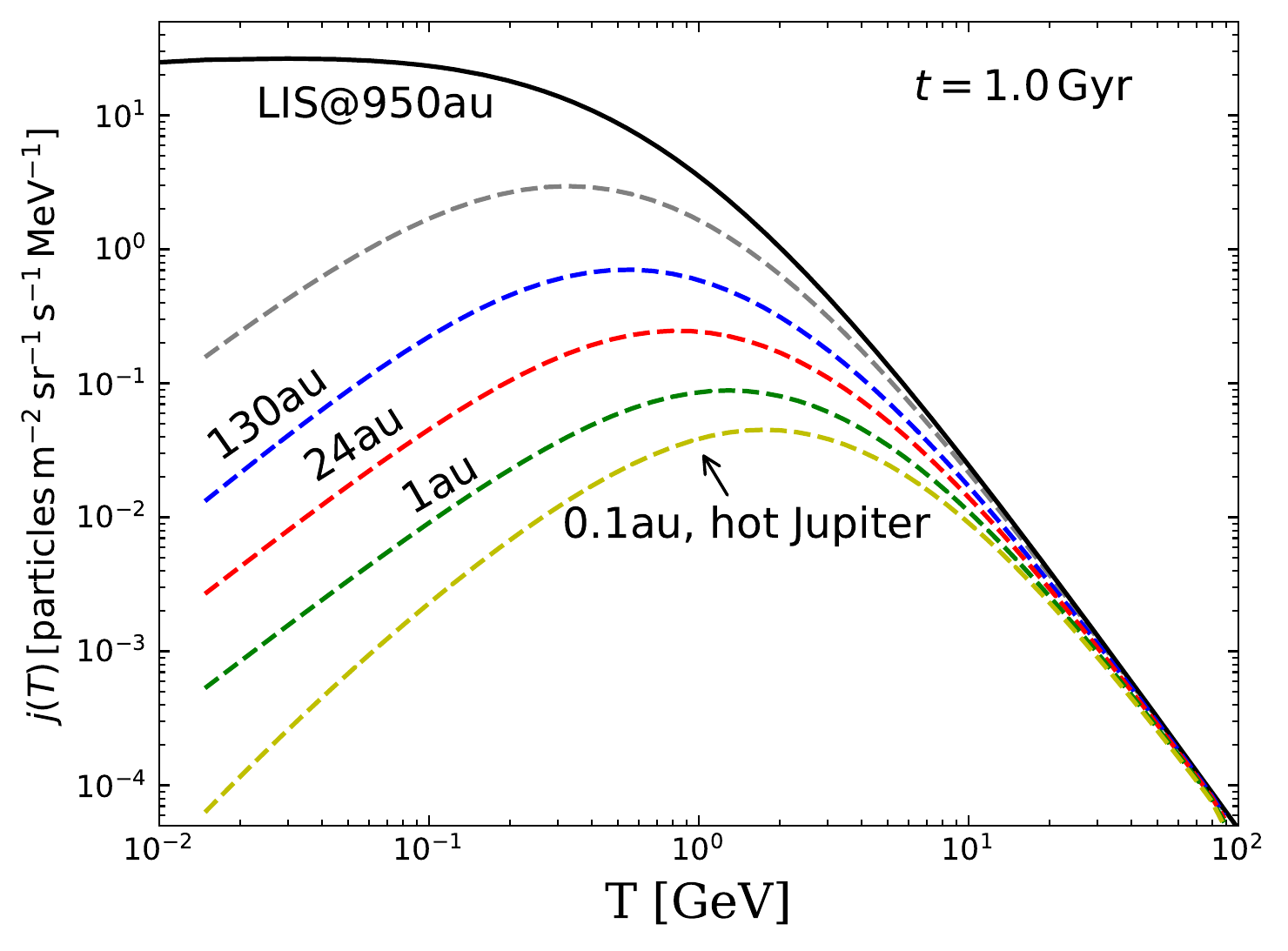}
       	\centering
  \caption{Differential intensity of Galactic cosmic rays as a function of kinetic energy at different radii for $t=1.0\,$Gyr, when life is thought to have begun on Earth. The solid black line represents a model fit of the \textit{Voyager 1} data for the LIS (Section \ref{subsec:lis}) located at 950\,au at this younger age. The coloured dashed lines represent the differential intensity found at different radii of interest in the simulation. In particular, the green dashed line corresponds to the values found at 1\,au. For comparison, the grey dashed line indicates the values found at the present-day Earth.} 
   \label{fig:gyr}
\end{figure}

Here we focus on the Galactic cosmic ray spectrum for a number of different radii at $t = 1.0\,$Gyr, shown in Fig.\,\ref{fig:gyr}, at approximately the time when life is thought to have begun on Earth. The first noticeable feature is that, because the heliosphere was much larger at this earlier time in the Sun's life (950\,au versus 122\,au), the differential intensity of cosmic rays at 130\,au (blue dashed line) is lower at most energies than the present-day values we observe at Earth (grey dashed line in Fig.\,\ref{fig:gyr}). We chose 130\,au, as this is approximately the present-day location of the edge of the heliosphere. The green dashed line denotes the values we find at 1\,au. For energies less than $\sim5\,$GeV these values are approximately 2 orders of magnitude smaller than the present-day values observed at Earth meaning that the young Earth was far better protected from Galactic cosmic rays than the present-day Earth. 

\section{Application to HR 2562b: Propagation of Galactic cosmic rays in the atmosphere of a young warm Jupiter}
\label{subsec:hr2562}
In the previous section we showed the Galactic cosmic ray spectrum that may have been present at the time when life began on Earth, or present at another Earth-like exoplanet orbiting at 1\,au from a young solar-type star. Observing the signatures of Galactic cosmic rays in the atmosphere of an Earth-like exoplanet would be important for understanding the origins of life on Earth and would also act as a constraint for the model we present. Unfortunately, it is unlikely with the current/near-future observing facilities that it would be possible to detect such a signature for an Earth-like exoplanetary atmosphere. Thus, in this section we focus on an exoplanetary system with a solar-type host star  where we believe it may be possible to detect the signatures of Galactic cosmic rays with the James Webb Space Telescope \citep{gardner_2006}. Our model can be used to guide future observations. It is important to note that even if the chemical effect of Galactic cosmic rays remains unobservable in Earth-like exoplanetary atmospheres that Galactic cosmic rays can still be important for these systems.

In order to detect an observable chemical effect driven by Galactic cosmic rays in an exoplanetary atmosphere with a solar-type host star we must first isolate the chemical effects of Galactic cosmic rays from other effects such as from photo-chemistry driven by stellar radiation or from stellar energetic particles. Stellar radiation and stellar energetic particles will generally dominate over Galactic cosmic rays in terms of observable signatures in the atmospheres of close-in exoplanets so we must focus on exoplanets at large orbital distances. Young exoplanets would also be easier to detect because exoplanets cool, and emit less flux, as they age. Thus, we apply our cosmic ray model to HR 2562, a young exoplanetary system with an estimated age of 300-900\,Myr \citep[see][for a discussion of the different age estimates for the star]{konopacky_2016}. This system hosts a warm (therefore meaning young) Jupiter exoplanet at a large distance from its host solar-type star -- HR 2562b is a directly imaged planet, observed as part of the Gemini Planet Imager Exoplanet Survey. 

HR 2562b has a mass of $30 \pm 15 M_\mathrm{Jup}$, orbiting a 1.3$M_\odot$ star (F5V) at a distance of $20.3\pm 0.3$\,au \citep{konopacky_2016}. At this orbital distance it is possible that Galactic cosmic rays will be more important than photo-driven chemistry in determining the chemical (dis-)equilibrium in the exoplanet's atmosphere.

\begin{figure}
	\centering
        \includegraphics[width=0.5\textwidth]{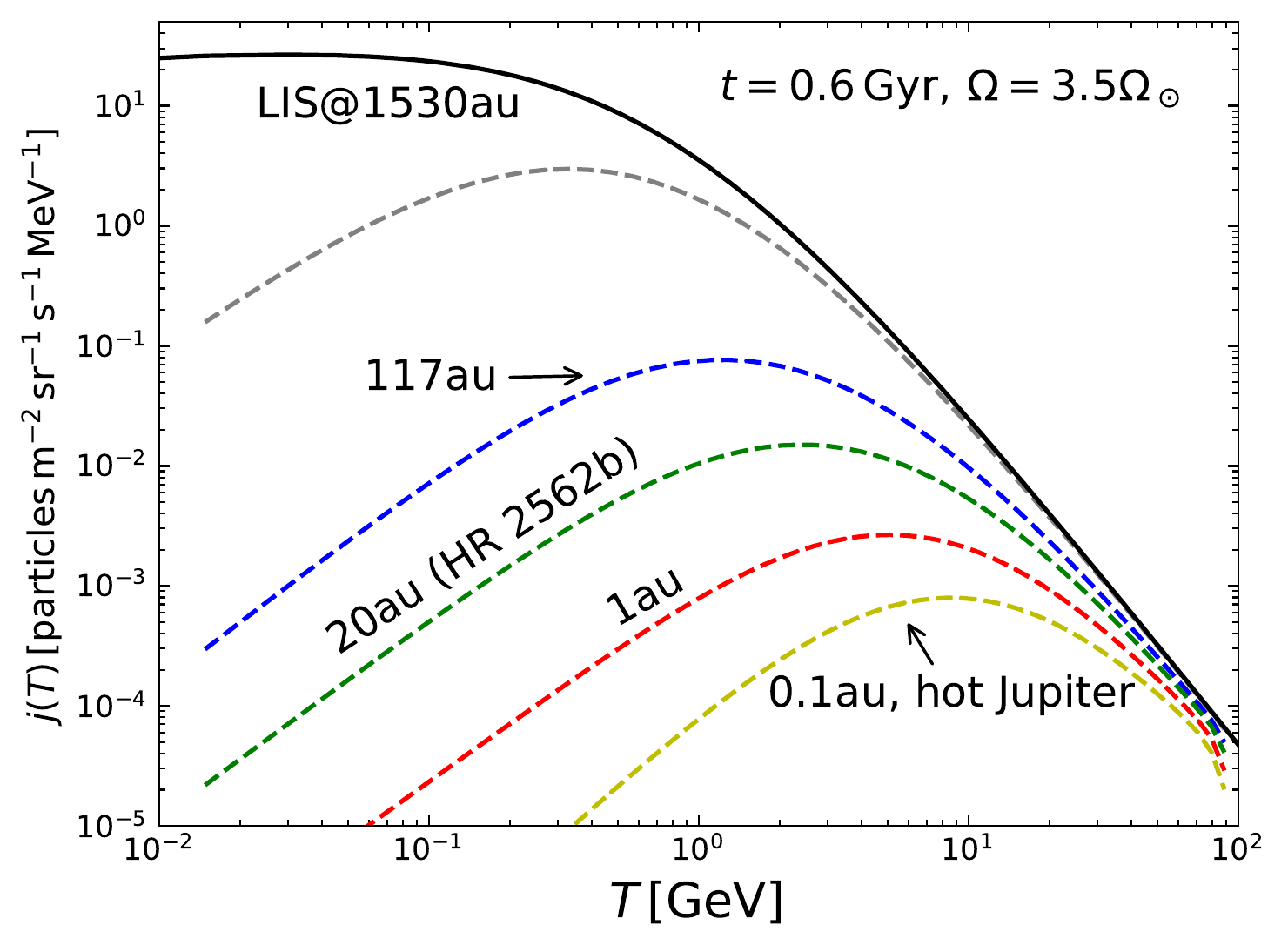}
       	\centering
  \caption{Differential intensity of Galactic cosmic rays as a function of kinetic energy at different radii for $t=0.6\,$Gyr. The solid black line represents a model fit of the \textit{Voyager 1} data for the LIS (discussed in \ref{subsec:lis}) located at 1530\,au. The coloured dashed lines represent the differential intensity found at different radii of interest in the simulation. In particular, the red dashed line corresponds to the values found at 1\,au and the green dashed line corresponds to the intensity of cosmic rays found at the same orbital distance as the exoplanet HR 2562b. For comparison, the grey dashed line indicates the values found at the present-day Earth.} 
   \label{fig:06gyr}
\end{figure}

To estimate the Galactic cosmic ray flux incident on HR 2562b, we use our Galactic cosmic ray spectrum for different radii at $t = 0.6\,$Gyr (using $\Omega = 3.5\Omega_\odot$). Fig.\,\ref{fig:06gyr} plots the differential intensity of Galactic cosmic rays at a number of different radii. The green dashed line corresponds to the orbital distance of the exoplanet HR 2562b.

We then use this Galactic cosmic ray spectrum to trace the subsequent propagation, and energy losses, of the cosmic rays down through the exoplanet's atmosphere using the Monte Carlo cosmic ray propagation model as described by \citet{rimmer_2013}. Here, we take into account energy losses due to inelastic (ionization and excitation) collisions (i.e. we neglect magnetic mirroring). \citet{rimmer_2013} contain further details of the Monte Carlo code. The atmosphere we use for our model is a DRIFT-PHOENIX atmosphere \citep{helling_2008a,helling_2008b,witte_2009} for a substellar object with an effective temperature $T_{\rm eff} = 1200$ K, surface gravity of $10^{4.5}$\,cm\,s$^{-2}$ and solar metallicity. 

\begin{figure}
	\centering
        \includegraphics[width=0.5\textwidth]{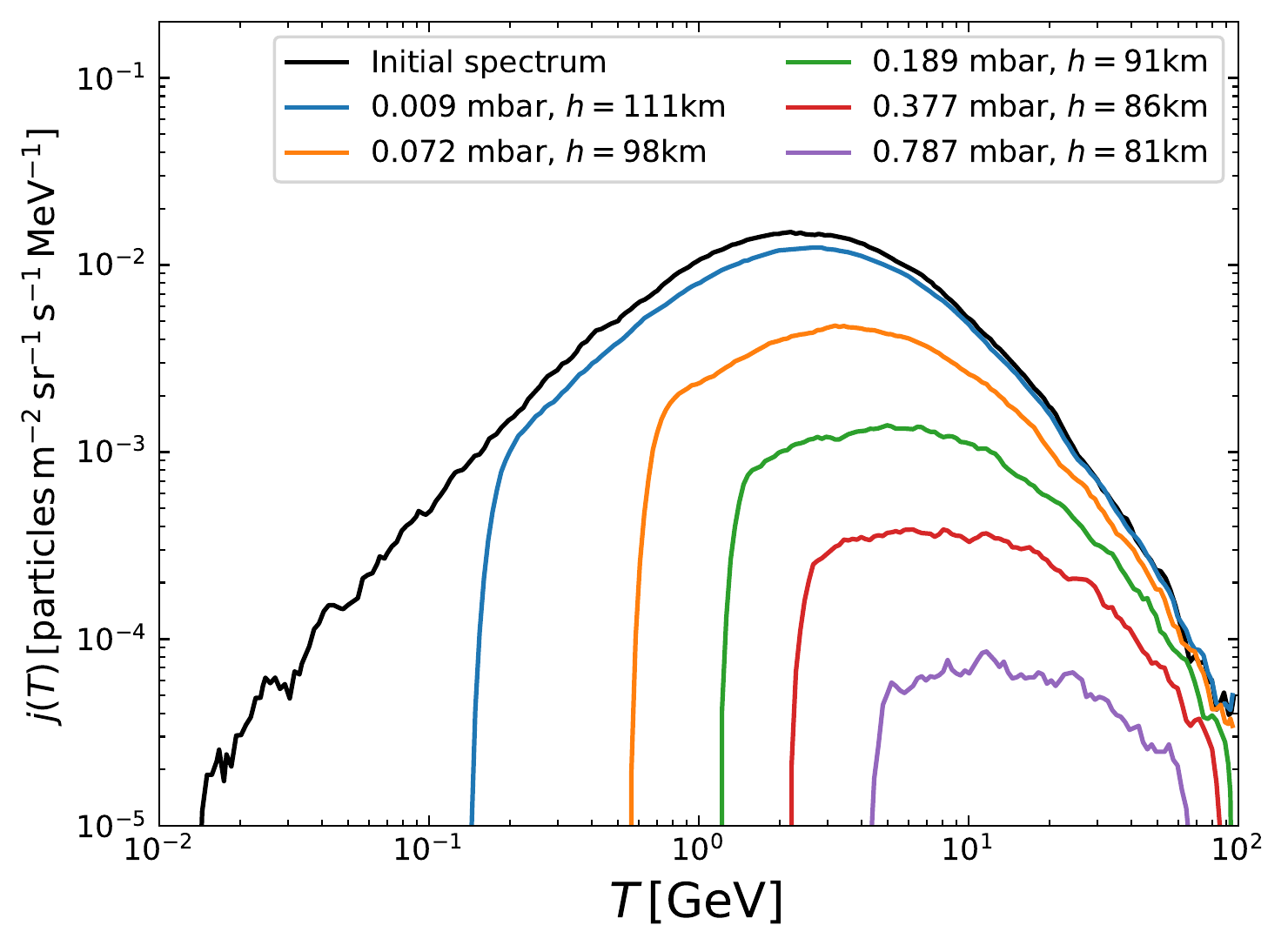}
       	\centering
  \caption{Differential intensity of Galactic cosmic rays as a function of kinetic energy for different atmospheric pressures, $P$, (and heights) within a model atmosphere for HR 2562b, based on a Monte Carlo cosmic ray propagation model \citep{rimmer_2013}.  }
   \label{fig:cr-spec}
\end{figure}

The resulting spectra for a range of atmospheric pressures, which correspond  to different atmospheric depths, are shown in Fig.\,\ref{fig:cr-spec}. The solid black line corresponds to the interpolation of the input spectrum (the green dashed line in Fig.\,\ref{fig:06gyr}) used to initialise the Monte Carlo code. The majority of particles $< 0.1$ GeV are attenuated at pressures greater than $10^{-5}$ bar, and the majority of $0.1$ -- $10$ GeV particles are attenuated at pressures greater than $10^{-4}$ bar. Much of the energy lost by cosmic rays will be deposited into the atmosphere by ionizing and dissociating various molecular species. This ionization and dissociation leads to the formation of the ions $\mathrm{H_3^+}$ and $\mathrm{H_3O^+}$ \citep{helling_2019}, and most of the formation will occur between 1 mbar and 1 bar. 
These species are rapidly destroyed by recombination with electrons, at a rate proportional to the pressure. The ions $\mathrm{H_3^+}$ and $\mathrm{H_3O^+}$ will be much more likely to survive at 1 mbar than 1 bar, and can then diffuse higher in the atmosphere. 
The results shown in Fig.\,\ref{fig:cr-spec} can be used to determine if the abundances of these molecules are observable using a chemical network model, such as the models presented in \citet{rimmer_2014,helling_2019} and \citet{moore_2019}, but is beyond the scope of this paper.

\section{Discussion: Comparison to the literature}
\label{sec:discussion}
Our simulation for $t=1.7$\,Gyr can be compared with the results of \citet{svensmark_2006}. The turquoise line in their Fig.\,1 corresponds to the same time denoted by the cyan dots in our Fig.\,\ref{fig:omega}. The peak flux occurs at approximately the same energy, i.e. $\sim$GeV. On the other hand the peak flux is approximately a factor of three larger in our simulation and at the lowest energies there is approximately one order of magnitude difference between the simulations. This difference at low energies is very likely due to differences in the adopted radial magnetic field and velocity profiles at small radii.

\citet{svensmark_2006} assumed a constant solar wind velocity as a function of radius and that the magnetic field scales as $r^{-1}$. In our case the solar wind velocity is only constant as a function of radius once it has reached its terminal velocity. The magnetic field scales as $r^{-1}$ beyond $r\sim1\,$au, whereas for $r<1\,$au it scales as $r^{-2}$ since the radial component of the magnetic field dominates at these radii. The evolution of the solar wind properties with time is also different between the two models which likely contributes to the differences seen between the two models. For $T\gtrsim$GeV using a constant solar wind velocity and $B\propto 1/r$ appears sufficient whereas at low energies it underestimates the differential intensity of Galactic cosmic rays.

Making a comparison with the results of \citet{cohen_2012} is less straightforward. We have used empirical relations from observations to estimate the temporal evolution of the solar wind properties as a function of the rotation rate as an input for our cosmic ray transport model. In contrast, \citet{cohen_2012} took an observed magnetic map of the Sun and  modified the map to mimic the presence of high latitude spots observed in young stars. Their Fig.\,4 represents the physical set-up most similar to our model where they have increased the dipole and spot component of the magnetic field by a factor of 10. The green line in their Fig.\,4 with a solar period of 10\,days is closest to our slow rotating Sun at $t=0.6\,$Gyr with $\Omega=3.0\Omega_\odot$. The peak intensity that they find is approximately a factor of 2 or 3 larger than our peak value. The kinetic energy at which the peak is found is very similar. 

While here we do not consider the interaction of the Galactic cosmic rays with an exoplanetary magnetic field \citep[as is the focus of][for a close-in exoplanet orbiting a M dwarf, for instance]{griessmeier_2015}, the differential intensity of Galactic cosmic rays that we find for different radii, and times in a solar-type star's life, can be used in future as an estimate for the boundary condition of simulations focusing on this interaction with exoplanets around other solar-type stars in more detail.

\section{Conclusions}
\label{sec:conclusions}
In this paper we investigated how the propagation of Galactic cosmic rays through the stellar systems' of solar-type stars would change as a function of the solar-type star's lifetime due to the varying physical conditions of the stellar wind with time. We modelled the modulation of Galactic cosmic rays by solving the associated 1D transport equation assuming diffusive transport, including spatial and momentum advection of Galactic cosmic rays by the stellar wind. We used a polytropic stellar wind model to derive the distance profile of the stellar wind for different stellar rotation rates.

We found that for a solar-type star older than the Sun ($t=6.0\,$Gyr) the differential intensity of Galactic cosmic rays will increase between a factor of 2-5 at $T \lesssim \,$GeV. At early ages, at $t=0.6\,$Gyr for instance,  the rotation rate of the Sun  is unknown. Therefore, we showed that the resulting difference in the differential intensity of Galactic cosmic rays at Earth, depending on whether the Sun was a fast or a slow rotator, is approximately an order of magnitude for $T \lesssim 5\,$GeV energies.

Generally, for mildly relativistic cosmic rays ($\lesssim$GeV energies) their associated diffusion timescales have always been comparable to, or longer than, the advective timescale of the stellar winds of solar-type stars. This means that the past and present modulation of these low energy cosmic rays in the solar system has always been severe. Only in the future, as the solar wind becomes weaker, will these low energy cosmic rays begin to reach Earth from the ISM. For faster rotation rates, approximately corresponding to younger ages, 10\,GeV cosmic rays begin to be severely modulated due to the increased magnetic field strength and velocity of the solar wind.

We compare our results to a modified version of the force field approximation and find that for rotation rates of $\Omega\lesssim 2.1 \Omega_\odot$ the modified force field approximation can be used to fit our results at 1\,au quite well. We provided an analytical fit to our derived spectra in Eq.\,\ref{eq:force-field}. These fits could be easily incorporated in future models, such as for calculating the spectrum at the top of Earth's atmosphere for the different ages that we focused on here.

We looked specifically at the differential intensity of Galactic cosmic rays that would have been incident on Earth at $t=1.0\,$Gyr, approximately when life is thought have begun on Earth. For $T\lesssim 5\,$GeV the values for the differential intensity that we find are approximately 2 orders of magnitude smaller than the present-day values observed at Earth, similar to previous estimates by \citet{cohen_2012}.

Finally, we applied our model to the case of HR 2562b which is a warm Jupiter orbiting a young $\sim$solar-like star ($t=0.6$\,Gyr) at 20\,au. After calculating the differential intensity of Galactic cosmic rays at the orbital distance of this exoplanet, we determine how the cosmic rays would deposit their energy as they propagate through the exoplanet's atmosphere. Here, we assumed the atmosphere to be unmagnetised. We found that the majority of cosmic ray particles with energies between 0.1 and 10 GeV are attenuated at pressures greater than $10^{-4}$\,bar. Our results can be used to guide future searches for the chemical signatures of Galactic cosmic rays in exoplanetary atmospheres with, for example, the JWST. An observational signature of Galactic cosmic rays in an exoplanetary atmosphere of a warm Jupiter may help constrain the Galactic cosmic ray spectrum present around young Earth-like exoplanets.

\section*{Acknowledgements}
The authors thank Dr Christiane Helling for providing the model atmosphere of HR 2562b. The authors also thank A. C. Cummings and B. Heikkila for providing the IMP 8 data. DRL and AAV acknowledge funding from the Irish Research Council Laureate Awards 2017/2018 and from the European Research Council (ERC) under the European Union's Horizon 2020 research and innovation programme (grant agreement No 817540, ASTROFLOW). P.~B.~R. thanks the Simons Foundation for support under SCOL awards 59963. This work has made use of DESY's high-performance computing facility. We would like to thank the anonymous referee for helpful comments which improved the manuscript.

\appendix
\section{The numerical code}
\label{appendix:numerics}
In this section we give details of the numerical code that was used including a description of the numerical scheme, how the boundary conditions are implemented, a definition of the overall timestep for the code as well as a validation of our code using Galactic cosmic ray observations at Earth and a resolution test. The code presented here assumes spherical symmetry and is adapted version of the code that was originally presented in \citet{rodgers-lee_2017} which had two spatial dimensions. The version of the code presented here uses a logarithmically spaced spatial grid \citep[which was used in][]{rodgers-lee_2020}, as well as a logarithmically spaced momentum grid which was not included in the previous version of the code. The last term in Eq.\,\ref{eq:f} describing momentum advection is also now included. We use a different numerical scheme for the advective terms which is described below.

\subsection{Numerical scheme}
Here we describe the numerical scheme used to discretise Eq.\,\ref{eq:f}. Both the spatial and momentum bins are logarithmically spaced and so we introduce a change of variables such that $u \equiv \ln r$ and $w \equiv \ln p$. Let $\tilde{\kappa}$ be the diffusion coefficient when written as a function of $u$ and $w$. Given any variable $X$, the notation $X^n_{i,j}$ denotes the variable $X$ at  $u_i$, $w_j$ and time $t^n$ with
\begin{equation}
u_i = i\Delta u, \hspace{5mm} w_j = j\Delta w, \hspace{5mm} t^n = n\Delta t
\end{equation}
where $\Delta u$ ($\Delta w$) is the radial (momentum) logarithmic grid spacing and $\Delta t$ is the timestep.

\noindent For the diffusive term in Eq.\,\ref{eq:f} we use a first order forward in time and second order centred in space scheme. The diffusion equation can be expressed in terms of $u$ as
\begin{eqnarray}
\frac{\partial f}{\partial t} &=&\bm{\nabla}\cdot(\kappa\bm{\nabla} f) \nonumber \\
&=& \frac{1}{r^2}\frac{\partial}{\partial r}\left( r^2 \kappa \frac{\partial f}{\partial r}
\right) \nonumber \\
&=& \frac{1}{r^3}\frac{\partial}{\partial u}\left( r\tilde{\kappa} 
\frac{\partial f}{\partial u}\right) = e^{-3u}\frac{\partial}{\partial u}\left( e^u\tilde{\kappa} 
\frac{\partial f}{\partial u}\right)
\end{eqnarray}
%\right|_{(i,j)}
We can discretise this using a forward in time, centred in space scheme as

\begin{eqnarray}
f^{n+1}_{i,j} = f^{n}_{i,j} + \frac{\Delta t}{(\Delta u)^2}  e^{-3u_i}\bigg( e^{-u_{i+1/2}}\tilde{\kappa}_{i+1/2,j}\left[f^n_{i+1,j}-f^n_{i,j}\right] \nonumber \\
 - e^{-u_{i-1/2}}\tilde{\kappa}_{i-1/2,j}\left[f^n_{i,j}-f^n_{i-1,j}\right] \bigg).
\end{eqnarray}
For the spatial advective term we use a finite volume first order in time and space upwinding scheme. Thus, written in conservative form the advection equation becomes
\begin{eqnarray}
\frac{\partial f}{\partial t} =  -\bm{\nabla}\cdot (\bm{v}f) =-\bm{\nabla}\cdot \bm{F}
\label{eq:advect}
\end{eqnarray}
with $\bm{F} = \bm{v}f$. Written in terms of $u$ this becomes,
\begin{equation}
\frac{\partial f}{\partial t} =  -e^{-3u} \frac{\partial}{\partial u}(e^{2u}vf). 
\label{eq:advect}
\end{equation}
Eq.\,\ref{eq:advect} can then be expressed as
\begin{equation}
\frac{\Delta f}{\Delta t} \approx -\left( \frac{F^n_{i+1/2,j}-F^n_{i-1/2,j}}{\Delta u}\right)   
\end{equation}
where $F_{i\pm 1/2,j}= e^{2u_{i\pm 1/2}}v^*_{i\pm 1/2}f^{n,*}_{i\pm1/2,j}$ with $v^*_{i\pm 1/2}$ and $f^{n,*}_{i\pm 1/2,j}$ being the so-called resolved states. Therefore
\begin{equation}
f^{n,*}_{i+1/2,j} = 
\begin{cases}
f^n_{i,j} \hspace{9mm}\text{if } v^*_{i+1/2}\geq0\\
f^n_{i+1,j}\hspace{6mm}		\text{if } v^*_{i+1/2}<0 \\
\end{cases}
\end{equation}
and similarly for $f^{n,*}_{i-1/2,j}$ where $v^*_{i+1/2} =(v_{i+1} +v_{i})/2$. Thus, written as a difference scheme this is
\begin{eqnarray}
f^{n+1}_{i,j}&=&f^{n}_{i,j} -\frac{\Delta te^{-3u_i}}{\Delta u}\bigg( \bigg.e^{2u_{i+1/2}}v^*_{i+1/2}f^{n,*}_{i+1/2,j} \nonumber \\
&-& e^{2u_{i-1/2}}v^*_{i-1/2}f^{n,*}_{i-1/2,j} \bigg. \bigg) .   
\end{eqnarray}
The momentum advection term is discretised in a similar way to the spatial advection term. Thus, the momentum advection term can be expressed as, 
\begin{equation}
\frac{\partial f}{\partial t} =  \frac{(\bm{\nabla}\cdot\bm{v})}{3}\frac{\partial f}{\partial \mathrm{ln}p} = \frac{(\bm{\nabla}\cdot\bm{v})}{3}\frac{\partial f}{\partial w} .
\label{eq:mom_advect}
\end{equation}
For the 1D spherical case this becomes
\begin{equation}
\frac{\partial f}{\partial t} =  \frac{1}{3r^2}\frac{\partial}{\partial r}(r^2v)\frac{\partial f}{\partial w} = \frac{e^{-3u}}{3}\left(\frac{\partial e^{2u}v}{\partial u}\right)\frac{\partial f}{\partial w}
\label{eq:mom_advect}
\end{equation}
where we can rewrite this as a differencing scheme in terms of an effective velocity, $v'_{i,j+1/2}$, as
\begin{equation}
f^{n+1}_{i,j} = f^{n}_{i,j} + \frac{\Delta te^{-3u_i}}{3\Delta w}v'_{i,j+1/2}\left(f^{n,*}_{i,j+1/2} - f^{n,*}_{i,j-1/2} \right)   
\end{equation}
where 
\begin{equation}
v'_{i,j+1/2} = \left(\frac{\partial e^{2u}v}{\partial u}\right)_{i,j} = \frac{e^{2u_{i+1/2}}v_{i+1/2}-e^{2u_{i-1/2}}v_{i-1/2}}{\Delta u}
\end{equation}
and is independent of the index $j$. Finally, 
\begin{equation}
f^{n,*}_{i,j+1/2} = 
\begin{cases}
f^n_{i,j} \hspace{10mm}\text{if } v'_{i,j+1/2}\geq0\\
f^n_{i,j+1}\hspace{7mm}		\text{if } v'_{i,j+1/2}<0 \\
\end{cases}
\end{equation}
and similarly for $f^{n,*}_{i,j-1/2}$. Thus the overall scheme for Eq.\,\ref{eq:f} is given by,
\begin{eqnarray}
f^{n+1}_{i,j} &=& f^{n}_{i,j} + \frac{\Delta t}{(\Delta u)^2}  e^{-3u_i}\bigg( e^{-u_{i+1/2}}\tilde{\kappa}_{i+1/2,j}\left[f^n_{i+1,j}-f^n_{i,j}\right] \bigg. \nonumber\\
 &-& e^{-u_{i-1/2}}\tilde{\kappa}_{i-1/2,j}\left[f^n_{i,j}-f^n_{i-1,j}\right] \bigg. \bigg)\nonumber\\
 &-&\frac{\Delta te^{-3u_i}}{\Delta u}\bigg(e^{2u_{i+1/2}}v^*_{i+1/2}f^{n,*}_{i+1/2,j} \bigg. \nonumber \\
&-& e^{2u_{i-1/2}}v^*_{i-1/2}f^{n,*}_{i-1/2,j} \bigg. \bigg) \nonumber \\
&+& \frac{\Delta te^{-3u_i}}{3\Delta w}v'_{i,j+1/2}\left(f^{n,*}_{i,j+1/2} - f^{n,*}_{i,j-1/2} \right)  
\end{eqnarray}

\subsection{Boundary conditions}
\label{appendix:bcs}
The inner radial boundary condition is reflective meaning the cosmic rays cannot enter/leave via this boundary. To implement this boundary condition in the code we treat the spatial diffusion and advection terms separately. For the spatial advective term the velocity of the solar wind in the boundary cell is set to be the opposite of the velocity of the solar wind in the cell beside the boundary, i.e. $v_0 = -v_1$ which ensures that the advective flux across the boundary is zero ($v^*_{1/2}f^{n,*}_{1/2,j}=0$). To implement a reflective boundary for the diffusion term we ensure that the diffusive flux across the boundary is zero, i.e. $\kappa_{1/2,j}\nabla f|_{1/2,j}=0$. Therefore, $f^n_{0,j}=f^n_{1,j}$.  

The outer radial boundary condition is a fixed boundary condition set to the LIS value in the radial boundary cell. This is implemented in the code by simply fixing the value of the boundary cell to the LIS value which is constant in time. Cosmic rays can enter/leave the spatial grid via the outer radial boundary condition but they do not decrease/increase the value of the boundary cell. 

The lower and upper momentum boundary conditions are both outflow. This means no momentum is advected onto the momentum grid via the momentum boundaries, but momentum may leave the computational domain via these boundaries which requires no change to the current upwind numerical scheme. To ensure that momentum is not advected onto the grid, for the lower momentum boundary this requires that if $v'_{i,1/2}\geq 0$ then $f^{n,*}_{i,1/2}=0$. Similarly for the upper momentum boundary, if $v'_{i,M-1/2}\leq 0$ then $f^{n,*}_{i,M-1/2}=0$.

\subsection{Timestep}
To define the timestep for our scheme we first define a Courant condition for each separate term in Eq.\,\ref{eq:f}. Thus, the diffusive timestep is defined as $\Delta t_\mathrm{dif} = \mathrm{min}((\Delta x_i)^2/4\kappa_{i,j}) $, the spatial advection timestep is defined as $\Delta t_\mathrm{adv} = \mathrm{min}(\Delta x_i/v_i)$ and the momentum advection timestep is defined as $\Delta t_\mathrm{mom} = \mathrm{min} (3\Delta x_i \mathrm{ln(}\Delta p_j)/v_i) $.

Then, the overall timestep for the scheme is defined as
\begin{equation}
\Delta t =  \left( \frac{1}{\alpha\Delta t_\mathrm{dif} }+\frac{1}{\Delta t_\mathrm{adv} } +\frac{1}{\Delta t_\mathrm{mom}}\right)^{-1}
\end{equation}
where $\alpha=1/6$ is chosen. Since the diffusion coefficient and the velocity profile of the solar wind remain constant at a given simulated epoch the timestep for the scheme also remains constant for a given simulation run.

\subsection{Model validation using present-day data}
\label{appendix:observations}
We use current observations of Galactic cosmic rays at Earth and in the local ISM to compare with and constrain our numerical model. The Earth observations consist of IMP 8 \citep{mcdonald_1998}, BESS \citep[from][]{shikaze_2007} and PAMELA \citep[from][]{adriani_2013} data spanning a number of years. The local ISM observations are taken from \textit{Voyager 1} \citep{cummings_2016}. Our model can be seen to fit the observations well. An average magnetic field strength of 1.3\,G is used at the wind base, which is derived from a large scale magnetic field map of the Sun, as an input for the stellar wind model. We note that the value of 1.3\,G agrees with the observed magnetic field strength of the dipolar component of the Sun averaged over solar cycles 21 to 23 \citep[see Fig. 1 in][]{johnstone_2015b}. Overall, the results from our model at 1\,au match the observations quite well, with small discrepancies that are most likely due to the use of a simple 1D model to model an intrinsically asymmetric system. These small discrepancies could also be related to the variation of cosmic rays due to the solar cycle, which are not accounted for in the present paper. 

\begin{figure}
	\centering
        \includegraphics[width=0.5\textwidth]{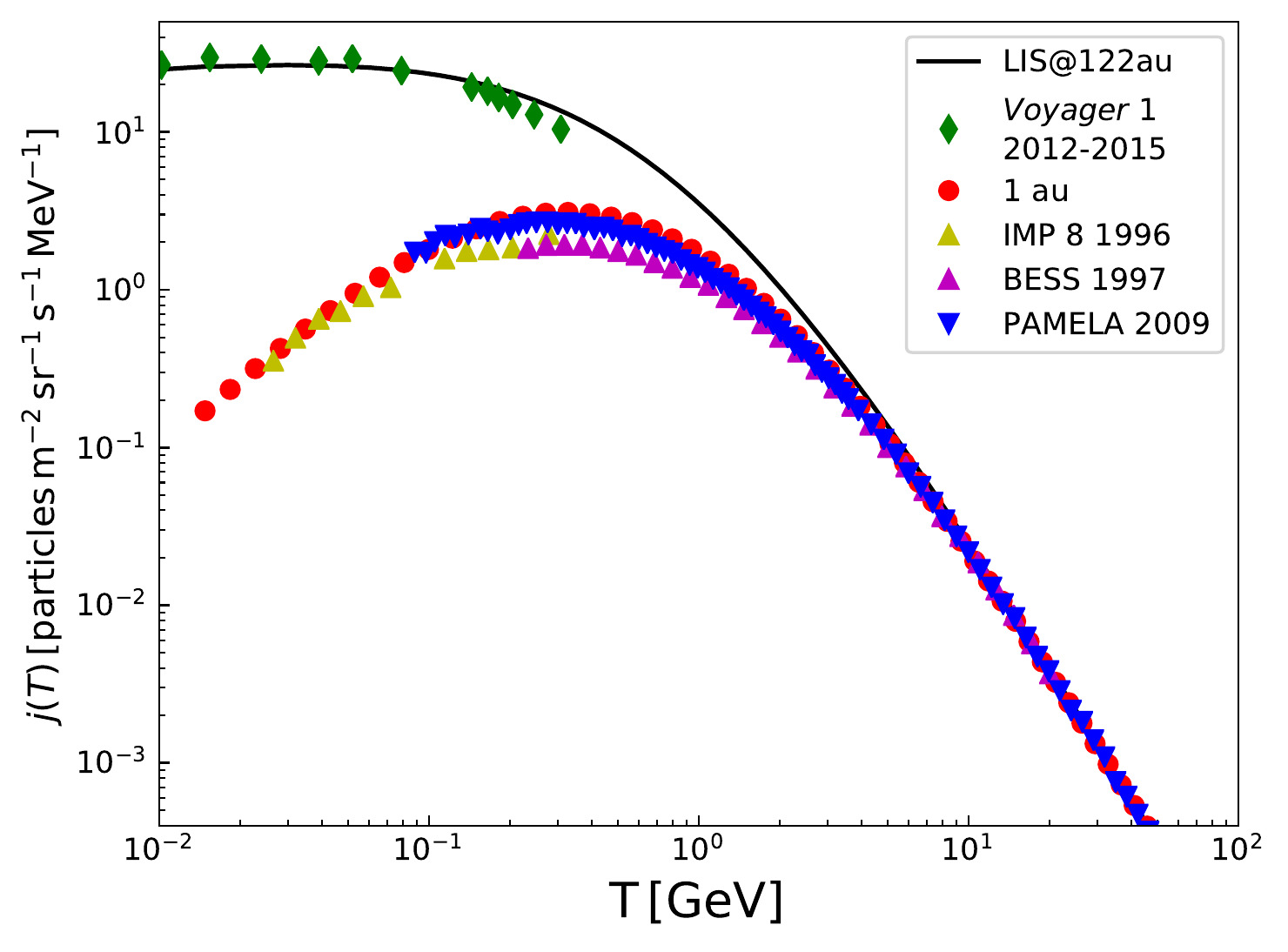}
       	\centering
  \caption{Differential intensity of Galactic cosmic rays as a function of kinetic energy. The solid black line represents a model fit of the \textit{Voyager 1} data (the green diamonds) for the LIS which are the values used at the spatial outer boundary (122\,au). The red dots represent our simulation results for 1\,au. The yellow, magenta and blue triangles are the IMP 8, BESS and PAMELA observations, respectively. } 
    \label{fig:standard}
\end{figure}

\subsection{Resolution test}
\label{appendix:resolution}

\begin{figure}
	\centering
        \includegraphics[width=0.5\textwidth]{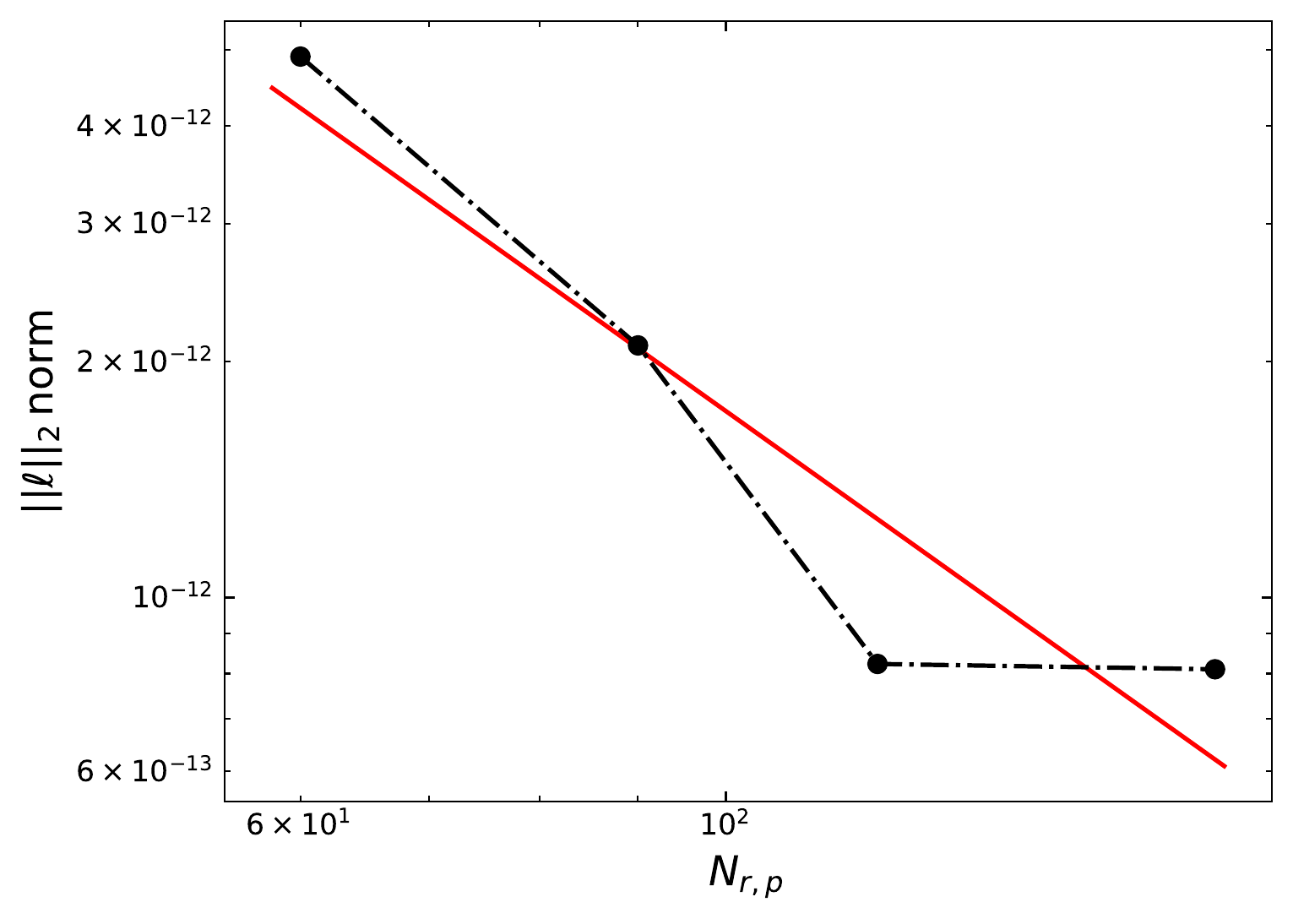}
       	\centering
  \caption[$||\ell||_2$ norm plotted as a function of resolution]{$||\ell||_2$ norm plotted as a function of resolution where $N_{r,p}$ means $N_r = N_p$ where $N_r$ is the number of grid zones in the spatial direction and $N_p$ is the number of momentum bins used.  } 
    \label{fig:l2-norm}
\end{figure}

We perform a resolution study using the $||\ell||_2$ norm for the simulation set-up using the present day values for the solar wind (given in Table\,\ref{table:sim_parameters}), shown in Fig.\,\ref{fig:l2-norm}. The $||\ell||_2$ norm is defined as
\begin{equation}
||\ell(a,b)||_2 = \sqrt{\frac{1}{n}\sum^n_{i=0}|x_{i,j;a}-x_{i,j;b}|^2}
\end{equation}

\noindent where the indices $i$ and $j$ indicate the spatial and momentum positions. The indices $a,b$ correspond to two simulations with different resolutions.  Five resolutions are considered increasing the number of bins in the radial (and momentum) direction with $N_\mathrm{r}(=N_\mathrm{p})=30,60,90,120,180$. The $||\ell||_2$ norm is calculated at the same time for each of the simulations. This time is chosen to be sufficiently large that the solution has effectively reached a steady state. A plot of $||\ell(a,b)||_2$ on a log-log scale should yield a straight line with a slope between -1 and -2 for our scheme since it is second order in space for the diffusive term but first order in space for advective terms. It is also first order in time but since the solutions are close to steady-state, as noted above, this will not manifest itself in this resolution study. The least-squares fitted slope of the data gives -1.74 indicating that the code is converging as expected and we conclude that our results are well resolved.

\section{Cosmic ray parameters}
\label{subsec:eta0_gamma}
 Throughout the paper we have used the same transport properties for the Galactic cosmic rays as a function of time. Here, we briefly discuss what this assumption physically implies about the system. The power law index, $\gamma$, from Eq.\,\ref{eq:kappa}, reflects the driving source of the turbulence in the solar wind which determines the turbulence power spectrum. The parameter $\eta_0$ describes the level of turbulence in the solar wind with a higher value meaning that the cosmic rays travel further before scattering.

Events such as coronal mass ejections (CMEs) are thought to drive of the turbulence in the solar wind but the exact connections still remain debated \citep{cranmer_2017}. Small scale convective motions on the solar surface \citep{mcintosh_2011} could additionally be transferred via Alfv{\'e}n waves to the large scale dipolar magnetic field structure and transported outwards in the solar wind but it is also possible that these waves will dissipate in the corona. Based on the solar flare-CME relation \citep{schmieder_2015} it is thought that young stars could produce more CMEs \citep{osten_2015} because they have been found to have higher flare rates \citep{maehara_2012}. This may lead to a stronger turbulent component of the magnetic field in the stellar system. At the same time young stars also have stronger magnetic fields and so how the ratio of $(B/\delta B)^2$ might change for a star younger than the Sun is overall unclear, as well as the fact that the stronger stellar magnetic fields of young stars may confine stellar CMEs \citep{alvarado-gomez_2018}. Generally though, a decrease in $(B/\delta B)^2$ means smaller diffusion coefficients which would increase the level of modulation suffered by Galactic cosmic rays. In our model we adopt $\eta_0=1$ which is already at the Bohm limit where the cosmic rays scatter once per gyroradius. Thus, in our model the magnetic field is already as turbulent as it can be using the diffusion approximation. If instead the level of turbulence in the magnetic field decreased as a function of increasing stellar rotation rate (i.e. larger values for $\eta_0$) the Galactic cosmic rays would not suffer as much modulation as presented here. For solar-type stars older than the Sun it is possible that a decrease in CME rates could result in less turbulence in the solar wind. This would lead to larger diffusion coefficients for Galactic cosmic rays and less modulation than is presented for $\Omega = 0.87\Omega_\odot$ in Fig.\,\ref{fig:omega}, for instance. 
\section{Magnetic field and velocity profiles from the stellar wind model}
\label{appendix:profiles}
\begin{figure*}%
	\centering
    \subfigure[ ]{%
        \includegraphics[width=0.5\textwidth]{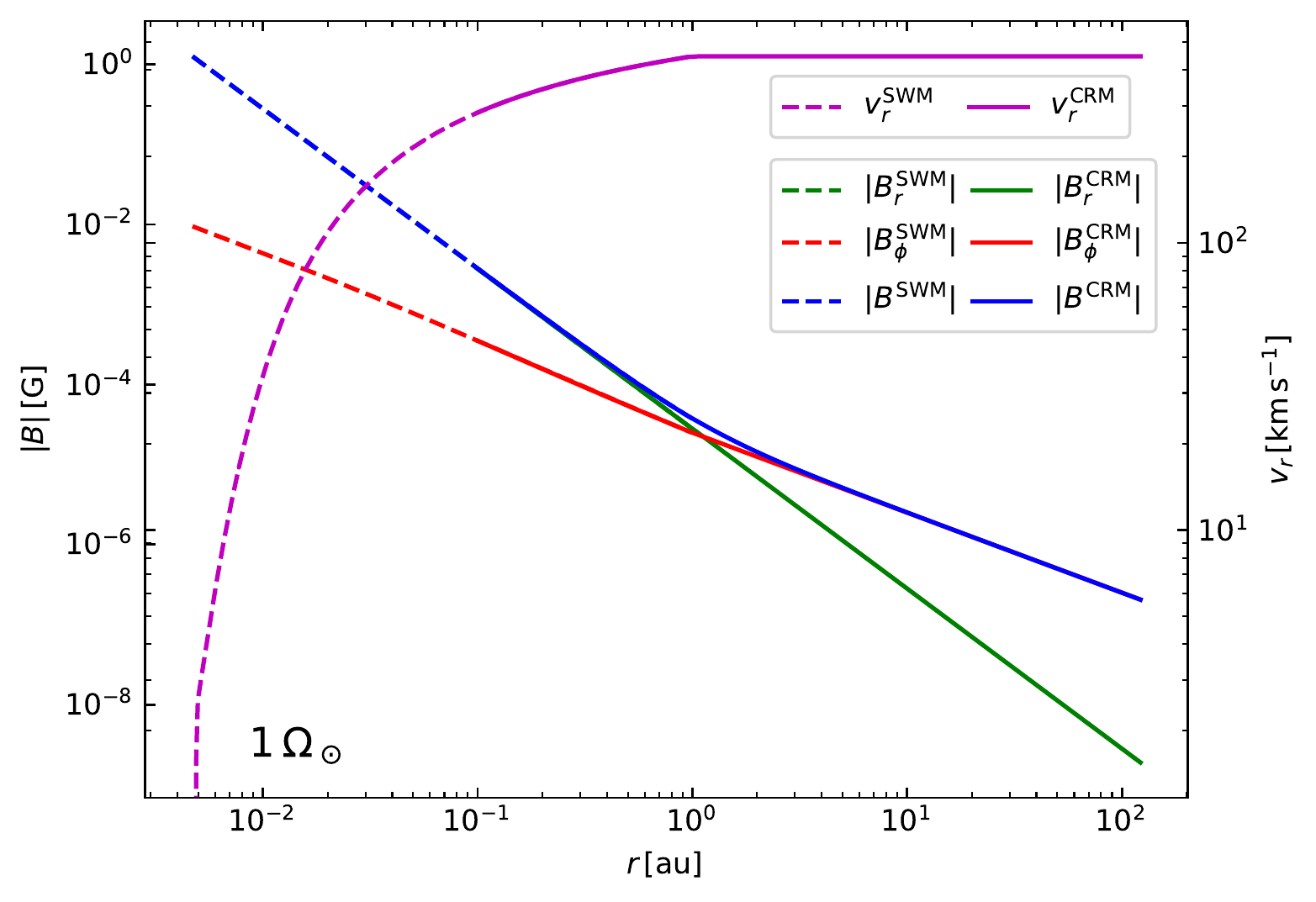}
       	\centering
\label{fig:omega1}}%
  	    ~
    \subfigure[ ]{%
        \includegraphics[width=0.5\textwidth]{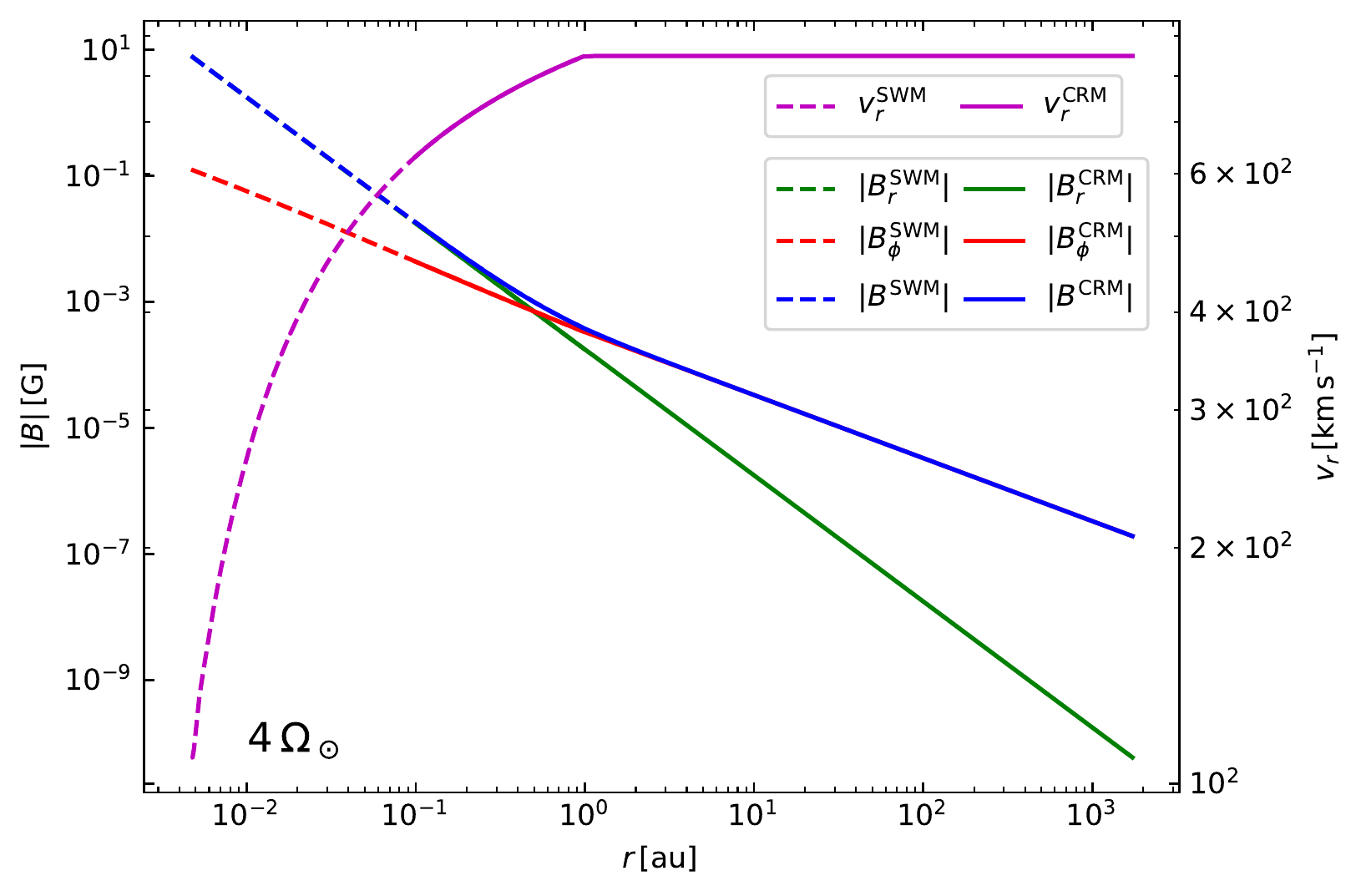}
	\centering
 \label{fig:omega4}}%
    \caption{(a) and (b) show the magnetic field and velocity profiles as a function of radius for two different stellar rotation rates, $1\Omega_\odot$ and $4\Omega_\odot$. The dashed lines represent the values derived from the stellar wind model (labelled as `SWM' in the plots) which extend to 1\,au. The solid lines represent the values used for the cosmic ray model (labelled as `CRM' in the plots) which assume the values from the stellar wind model out to 1\,au and then extrapolate the values from 1\,au out to the edge of the astrosphere as described in Section\,\ref{subsubsec:vel_mag}.} 
    \label{fig:profiles}%
\end{figure*}

In Fig.\,\ref{fig:profiles} we show the magnitude of the magnetic field components and the radial velocity as a function of radius for two of the rotation rates that we adopt ($1\Omega_\odot$ and $4\Omega_\odot$). The dashed lines represent values derived from the stellar wind model, as described in Section\,\ref{subsec:parameters_time} which extend to 1\,au. The solid lines represent the values that we use in the cosmic ray model which extend from 0.1\,au out to the edge of the stellar astrosphere. From 0.1-1\,au we use the values from the stellar wind model and beyond 1\,au we extrapolate from the values of the quantities at 1\,au as described in Section\,\ref{subsubsec:vel_mag}.

\section{Modified force field approximation comparison}
\label{appendix:ffa}
\begin{figure}
	\centering
        \includegraphics[width=0.5\textwidth]{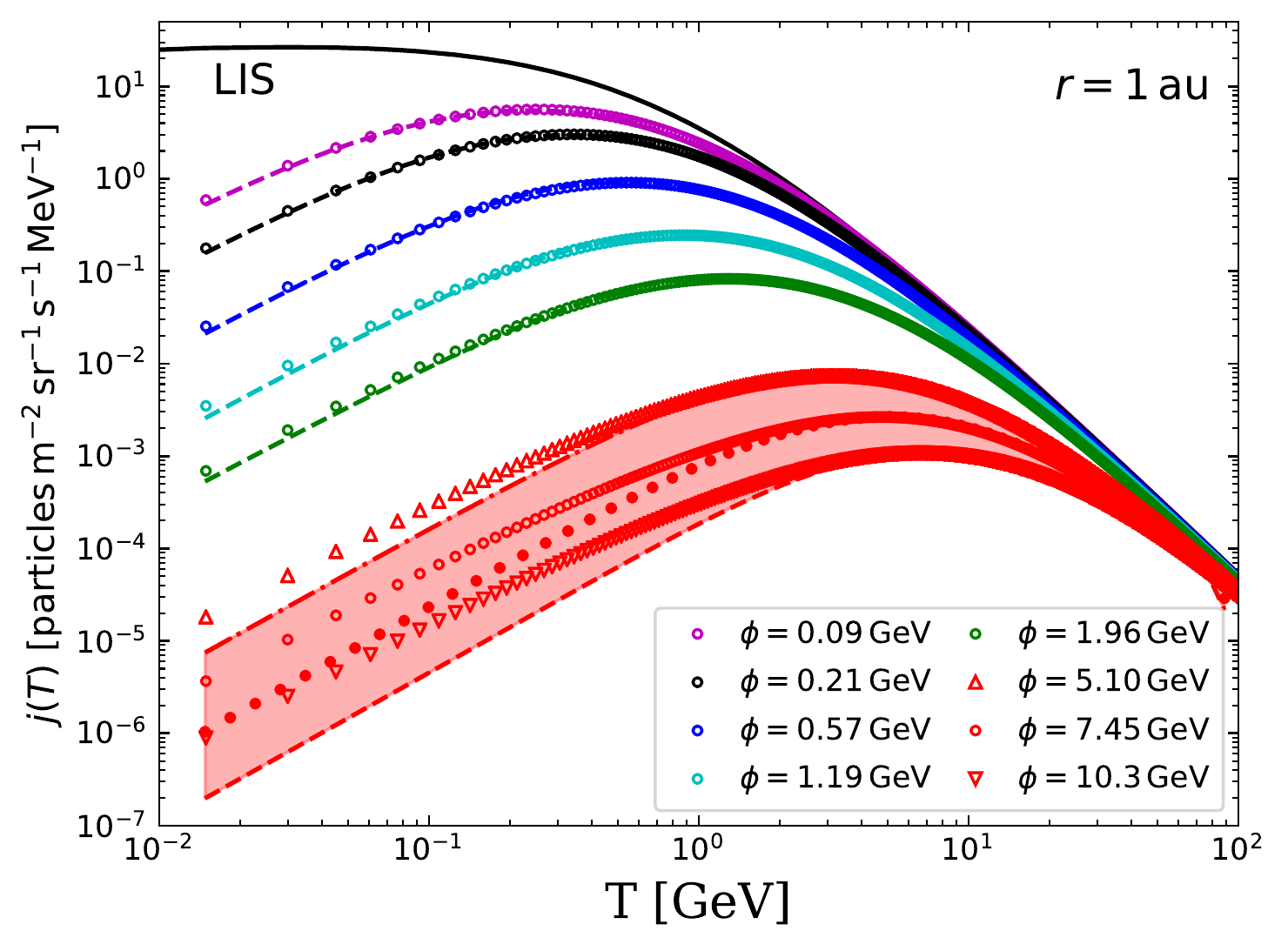}
       	\centering
  \caption{Differential intensity of Galactic cosmic rays as a function of kinetic energy. The solid black line represents a model fit of the \textit{Voyager 1} data for the LIS. The coloured dashed lines and the red shaded area represent our simulation results, as shown in Fig.\,\ref{fig:omega}. The open symbols represent the modified force field approximation which fits our simulation results well, with the value of $\phi$ that is used shown in the figure. See Section\,\ref{subsubsec:force-field} for more details. } 
    \label{fig:ffa_comparison}
\end{figure}

Here, in Fig.\,\ref{fig:ffa_comparison}, we present the comparison of our simulation results with the modified force field approximation. Our simulations results showed more suppression at low energies than the normal force field approximation. This led us to provide a modified force field approximation,  given in Eq.\,\ref{eq:force-field}, which matches our results at 1\,au well for $\Omega\leq 2.1 \Omega_\odot$. Therefore Eq.\,\ref{eq:force-field}, along with the values of $\phi$ given in Table\,\ref{table:sim_parameters}, can be used to reproduce these results. For $\Omega > 2.1 \Omega_\odot$, the modified force field approximation matches the peak well but fails to reproduce our simulation results at the lowest kinetic energies. 

\section*{Data Availability}
The data underlying this article will be shared on reasonable request to the corresponding author.

\newcommand\aj{AJ} %Astronomical Journal
\newcommand\actaa{AcA} %Acta Astronomica
\newcommand\araa{ARA\&A} %Annual Review of Astron and Astrophys
\newcommand\apj{ApJ} %Astrophysical Journal
\newcommand\apjl{ApJ} %Astrophysical Journal, Letters
\newcommand\apjs{ApJS} %Astrophysical Journal, Supplement
\newcommand\aap{A\&A} %Astronomy and Astrophysics
\newcommand\aapr{A\&A~Rev.} %Astronomy and Astrophysics Reviews
\newcommand\aaps{A\&AS} %Astronomy and Astrophysics, Supplement
\newcommand\mnras{MNRAS} %Monthly Notices of the RAS
\newcommand\pasa{PASA} %Publications of the Astron. Soc. of Australia
\newcommand\pasp{PASP} %Publications of the ASP
\newcommand\pasj{PASJ} %Publications of the ASJ
\newcommand\solphys{Sol.~Phys.} %Solar Physics
\newcommand\nat{Nature} %Nature
\newcommand\bain{Bulletin of the Astronomical Institutes of the Netherlands}
\newcommand\memsai{Mem. Societa Astronomica Italiana}
\newcommand\apss{Ap\&SS} % Astrophysics and Space Science
\newcommand\qjras{QJRAS} % Quarterly Journal of the RAS
\newcommand\pof{Physics of Fluids}
\newcommand\grl{Geophysical Research Letters}
\newcommand\planss{Planetary and Space Science}
\newcommand\ssr{Space Science Reviews}
\newcommand\astrobiology{Astrobiology}
\newcommand\icarus{Icarus}

\bibliographystyle{mn2e}
\bibliography{donnabib}

\label{lastpage}

\end{document}